\documentclass{elsart}

\usepackage{natbib}

\usepackage{graphicx}

\usepackage{amssymb}

\usepackage{subfig}

\def\s{$\sim$}
\def\sec{\section}

\def\bea{\begin{equation}}
\def\eea{\end{equation}}
\def\ber{\begin{eqnarray}}
\def\eer{\end{eqnarray}}

\def\araa{ARA\&A, }             
\def\apj{ApJ, }                 
\def\apjl{ApJ, }                
\def\apjs{ApJS, }               
\def\aap{A\&A, }                
\def\mnras{MNRAS, }             
\def\nat{Nature, }              

\makeatletter
\DeclareRobustCommand*\textsubscript[1]{%
  \@textsubscript{\selectfont#1}}
\def\@textsubscript#1{%
  {\m@th\ensuremath{_{\mbox{\fontsize\sf@size\z@#1}}}}}
\makeatother
\newcommand\tsb{\textsubscript}

\begin{document}

\begin{frontmatter}



\title{Refining a relativistic, hydrodynamic solver: Admitting ultra-relativistic flows}


\author{J.~P. Bernstein}
\address{Argonne National Laboratory, High Energy Physics Division, Argonne, IL 60439 and 
Department of Astronomy, University of Michigan,
500 Church Street, 830 Dennison, Ann Arbor, MI 48109-1042}
\ead{jpbernst@umich.edu}

\author{P.~A. Hughes}
\address{Department of Astronomy, University of Michigan,
500 Church Street, 830 Dennison, Ann Arbor, MI 48109-1042}
\ead{phughes@umich.edu}

\begin{abstract}
We have undertaken the simulation of hydrodynamic flows with bulk
Lorentz factors in the range 10$^2$--10$^6$. We discuss the
application of an existing relativistic, hydrodynamic
primitive-variable recovery algorithm to a study of pulsar winds,
and, in particular, the refinement made to admit such
ultra-relativistic flows. We show that an iterative
quartic root finder breaks down for Lorentz factors above 10$^2$ and employ
an analytic root finder as a solution. We find that the former, which is known to be
robust for Lorentz factors up to at least 50, offers a 24\% speed
advantage. We demonstrate the existence of a simple diagnostic
allowing for a hybrid primitives recovery algorithm that includes an
automatic, real-time toggle between the iterative and analytical
methods. We further determine the accuracy of the iterative and
hybrid algorithms for a comprehensive selection of input parameters
and demonstrate the latter's capability to elucidate the internal structure of 
ultra-relativistic plasmas. In particular, we discuss simulations 
showing that the interaction of a light, ultra-relativistic pulsar wind with
a slow, dense ambient medium can give rise to asymmetry reminiscent of the 
Guitar nebula leading to the formation of a relativistic backflow harboring 
a series of internal shockwaves. The shockwaves provide thermalized energy 
that is available for the continued inflation of the PWN bubble. In turn, 
the bubble enhances the asymmetry, thereby providing positive feedback to the backflow.
\end{abstract}


\end{frontmatter}
\sec{Introduction}
Hydrodynamic simulations have been widely used to model a broad range of physical systems.
When the velocities involved are a small fraction of the speed of light and gravity is weak,
the classical Newtonian approximation to the equations of motion may be used. However, these
two conditions are violated for a host of interesting scenarios, including, for example, heavy ion collision systems \citep{hir04},
relativistic laser systems \citep{del05}, and many from astrophysics \citep[][and references therein]{iba03}, that call for
a fully relativistic, hydrodynamic (RHD) treatment. The methods
of solution of classical hydrodynamic problems have been successfully adapted to those of a RHD
nature, albeit giving rise to significant complication; in particular, the physical quantities of a hydrodynamic flow
(the rest-frame mass density, $n$, pressure, $p$, and
velocity, $v$) are coupled to the conserved quantities (the laboratory-frame mass density, $R$, momentum density, $M$, and energy density, $E$)
via the Lorentz transformation. The fact that modern RHD codes typically evolve
the conserved quantities necessitates the recovery of the physical quantities (often referred to as the ``primitive variables'')
from the conserved quantities in order to obtain the flow velocity. Thus, the calculation of the primitives from the
conserved variables has become a critical element of modern RHD codes \citep{mar03}. Indeed, this is an active area of research with
significant attention given to the case of general relativistic, magnetohydrodynamic (GRMHD) case \citep[e.g.][]{nob06} and varying
equations of state within the context of RMHD \citep[e.g.][]{mig07} and RHD \citep[e.g.][]{ryu06}. This work is concerned with the RHD case for a fixed adiabatic index and so we refer the reader to the above-mentioned papers for a discussion of those studies.

In this paper, we present a method for recovering the primitive variables from the conserved quantities
representing special relativistic, hydrodynamic (SRHD) flows with bulk Lorentz factors ($\gamma =
(1-v^2)^{-1/2}$, where $v$ is the bulk flow velocity -- the speed of light is normalized to unity throughout this paper) up to 10$^6$. We started with a module
from an existing SRHD code used to simulate flows with $\gamma \le$50 as described in \cite{dun94}. Admitting flows with such
ultra-relativistic Lorentz factors as 10$^6$ required significant refinement to the method used in the existing code to calculate the
flow velocity from the conserved quantities. In particular, such extreme Lorentz factors lead to severe numerical problems such as effectively dividing by zero and subtractive cancellation. In $\S$\ref{hydro}, we discuss the formalism of recovering the primitives within the context of the Euler equations. In $\S$\ref{sfix}, we elucidate the details of the refinement
to this formalism necessitated by ultra-relativistic flows. We present the refined primitives algorithm in $\S$\ref{solver} and our application in $\S$\ref{appl}.
We discuss our results and conclusions in $\S$\ref{sec:dis} and $\S$\ref{sec:conc}, respectively.

\sec{Recovering the primitive variables from $R$, $M$, and $E$}{\label{hydro}}
In general, recovering the primitives from the conserved quantities reduces to solving a quartic equation, $Q(v)=0$, for the flow velocity in terms of $R$, $M$, and $E$. Implementation typically involves a numerical root finder to recover the velocity via Newton-Raphson iteration which is very efficient and provides robustness because it is straightforward to ensure that the computed velocity is always less than the speed of light. This is a powerful method that is independent of dimensionality and symmetry. The latter point follows directly from the fact that symmetry is manifest only as a source term in the Euler equations  and does not enter into the derivation of $Q(v)$ (see the axisymmetric example below). Dimensional generality arises because regardless of the coordinate system, one may always write $M=\sqrt{\sum M_{x_i}^2}$, where the $M_{x_i}$ are the components of the momentum-density vector along the orthogonal coordinates $x_i$. In the case of magnetohydrodynamic (MHD) flows, there are, of course, additional considerations. However, non-magnetic (RHD) simulations still have a significant role to play in astrophysics, e.g. extragalactic jets \citep{hug05} and pulsar wind nebulae \citep{swa04}.

As an example, consider the case of the axisymmetric, relativistic Euler equations, which we apply to pulsar winds. This type of formalism enjoys diverse application, in both special and general relativistic settings, from 3D simulations of extragalactic jets \citep{hug02}, to theories of the generation of gamma-ray bursts \citep{zha03} and the collapse of massive stars to neutron stars and black holes \citep{shi03}. In cylindrical coordinates $\rho$ and $z$, and defining the evolved-variable, flux, and source vectors

\ber
U & = & (R,M_{\rho},M_{z},E)^{T}\rm,\nonumber\\
F^{\rho} & = & (Rv^{\rho},M_{\rho} v^{\rho} + p, M_{z} v^{\rho},(E + p)v^{\rho})^T\rm,\nonumber\\
F^z & = & (Rv^{z},M_{\rho} v^{z},M_{z} v^{z} + p,(E + p)v^z)^T\rm,\nonumber\\
S & = & (0,p/{\rho},0,0)^{T}\rm,
\label{vecs}
\eer
the Euler equations may be written in almost-conservative form as:

\ber
{\partial{ U}\over \partial{t}} + {1\over\rho}{\partial\over \partial{\rho}}
(\rho F^{\rho}) + {\partial\over\partial z} (F^{z}) & = &S\rm.\nonumber
\eer
The pressure is given by the ideal gas equation of state
$p = (\Gamma - 1)(e - n)$,
where $e$ and $\Gamma$ are the rest-frame total energy density
and the adiabatic index. Note that the velocity and pressure appear
explicitly in the relativistic Euler equations, in addition to the evolved
variables, and pressure and rest density are needed for the computation of
the wave speeds that form the basis of typical numerical hydrodynamic solvers,
such as that due to \cite{god59}. We obtain
these values by performing a Lorentz transformation where the rest-frame
values are required:

\ber
R & = & \gamma n\rm,\nonumber\\
M_{\rho} & = & \gamma^2 ( e + p ) v^{\rho}\rm,\nonumber\\
M_{z} & = & \gamma^2 ( e + p )v^{z}\rm,\nonumber\\
E & = & \gamma^2 ( e + p ) - p \rm,\nonumber\\
\gamma & = & (1-v^2)^{-1/2}\rm,
\label{lt}
\eer
where $v^2 = (v^{\rho})^2 + (v^{z})^2$ and $M^2$ = $\gamma^4(e+p)^2[(v^{\rho})^2 + (v^{z})^2
]$ = $\gamma^4(e+p)^2v^2$. When the adiabatic index is constant, combining the above
equations with the equation of state creates a closed system which yields the following
quartic equation for $v$ in terms of $Y\equiv M/E$ and $Z\equiv R/E$:

\ber
Q(v) & = & (\Gamma-1)^2(Y^2+Z^2)v^4 - 2\Gamma(\Gamma-1)Yv^3\nonumber\\
& + & \Bigl\lbrack\Gamma^2+2(\Gamma-1)Y^2-(\Gamma-1)^2Z^2\Bigr\rbrack v^2\nonumber\\
& - & 2\Gamma Yv + Y^2\;=\;0\rm.
\label{qv}
\eer
Component velocities, and the rest-frame total energy and mass densities are
then given by:

\ber
v^{\rho} & = & {M_{\rho}\over M} v\rm,\nonumber\\
v^{z} & = & {M_{z}\over M_{\rho}}v^{\rho} \rm,\nonumber\\
e & = & E - M_{\rho}v^{\rho} - M_{z} v^{z}\rm,\nonumber\\
n & = & {R\over \gamma} \rm.\nonumber
\eer

\sec{Refinement of the root finder to admit ultra-relativistic flows}\label{sfix}
A particular implementation of the above has been previously applied to relativistic galactic
jets with $\gamma \leq 50$ \citep{dun94}. The ultra-relativistic nature of pulsar winds
necessitated an investigation of the behavior of the primitives algorithm
upon taking $\gamma >> 1$. We found that, beyond $\gamma \sim 10^2$, the
algorithm suffers a severe degradation in accuracy that worsens with increasing Lorentz
factor until complete breakdown occurs due to the failure of the Newton-Raphson
iteration process used to calculate the flow velocity.

\begin{figure}[h]
\caption{The left-hand plots show the shape of the Lorentz factor
quartic over a run of Lorentz factors for a mildly relativistic flow
($\gamma_o = 1.5$) and an ultra-relativistic flow ($\gamma_o =
10^6$). The right-side plots show the shape of the velocity quartic
over a run of velocity for a mildly relativistic flow ($\gamma_o
\approx 1.5$) and a highly (but not ultra) relativistic flow
($\gamma_o \approx 10^2$). The crosses mark the location of the
physical root. From the plot in the lower right, one can see the
onset of the zero derivative problem as the roots are not
distinguishable from each other or the local minimum even on a scale
of 10$^{-13}$, which begins to encroach on the limit of 8-byte
accuracy.}
\centerline{\includegraphics[width=4in]{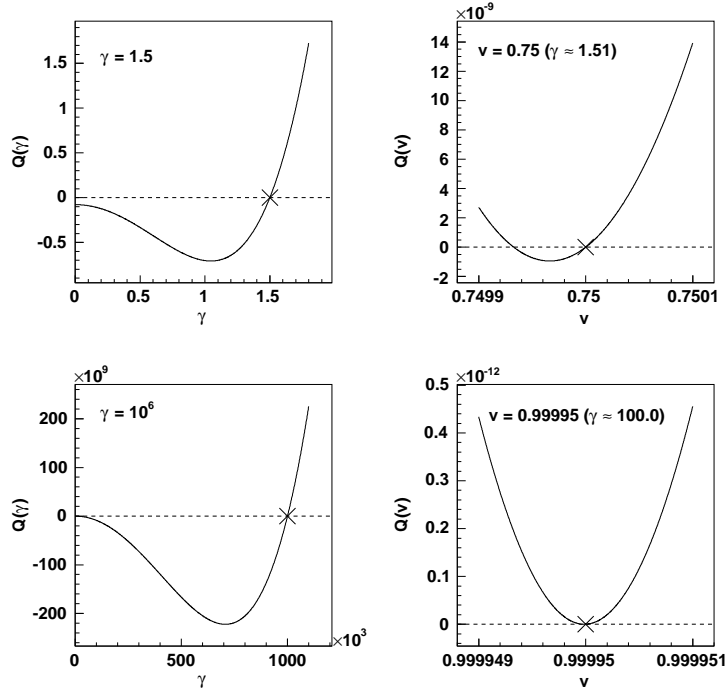}}
\label{lf_fix}
\end{figure}

The problem lies in the shape of the quartic, $Q(v)$, one must solve
to calculate the primitive variables. The quartic equation as derived using
the velocity as a variable exhibits two roots for typical
physical parameters of the flow (see Fig.~\ref{lf_fix}). In general,
for $\gamma < 10^2$, the two roots are sufficiently separated on the velocity axis
such that the Newton-Raphson (N-R) iteration method converges to the correct zero very
quickly and accurately (for $Y < 0.9$ and $Z > 10^{-5}$, corresponding to $\gamma < 2$, the roots approach each other sufficiently such that the incorrect root is selected; see $\S$\ref{solacc}). In fact, N-R iteration can be so efficient that it is
more desirable to use this method than it is to calculate the roots of the quartic
analytically (see $\S$\ref{timing}). However, as the Lorentz factor of the flow increases, the
roots move progressively closer together and the minimum in $Q(v)$ approaches zero. Eventually,
the minimum equals zero to machine accuracy which causes $dQ/dv = 0$ to machine accuracy
resulting in a divide by zero and the Newton-Raphson method fails (see Fig.~\ref{nr_acc}).

A simple and highly effective solution (see $\S$\ref{solacc} for details) is to rewrite the velocity quartic,
$Q(v)$ (Eqn.~\ref{qv}), in terms of the Lorentz factor (i.e. make the
substitution $v^2 = 1 - \gamma^{-2}$) to obtain the quartic equation in $\gamma$ (recall $Y\equiv M/E$ and $Z\equiv R/E$):

\ber
Q(\gamma) & = & \Gamma^2(1-Y^2)\gamma^4 - 2\Gamma(\Gamma-1)Z\gamma^3\nonumber\\
& + & \Bigl\lbrack2\Gamma(\Gamma-1)Y^2+(\Gamma-1)^2Z^2-\Gamma^2\Bigr\rbrack\gamma^2\nonumber\\
& + & 2\Gamma(\Gamma-1)Z\gamma - (\Gamma-1)^2(Y^2+Z^2)\;=\;0\rm.
\label{qg}
\eer
As Fig.~\ref{lf_fix} exemplifies, $Q(\gamma)$ exhibits a single root
for the physical range $\gamma \ge$ 1. However, Newton-Raphson iteration
also fails in this case at high Lorentz factors because of the steepness of the
rise in $Q(\gamma)$ through the root. Thus, we are forced to use an analytical
method of solving a quartic. Below, we discuss our implementation.

\subsection{Solving a quartic equation}\label{solveqe}
We use the prescription due to \cite{bro97} in order to analytically solve for the roots of a quartic.
We chose this method because it provides equations for
the roots of the quartic that are the most amenable (of the methods surveyed)
to integration into a computational environment. In order to provide a complete picture
of our method, which includes steps not found in \cite{bro97}, we reproduce
some sections of that text. We procede as follows.

Given a quartic equation in $x$:

\bea
a_4x^4+a_3x^3+a_2x^2+a_1x+a_0 = 0,\;a_n\in\Re,\;a_4\neq0\rm,
\label{qrt}
\eea
normalizing the equation (dividing by $a_4$) and making the substitution
$y = x + \frac{a_3}{4a_4}$ results in the reduced form:

\ber
y^4+Py^2+Qy+R & = & 0\nonumber\rm,
\eer
where, defining $\tilde{a}_n \equiv a_n/a_4$:

\ber
P&\equiv&-\frac{3}{8}\tilde{a}_3^2+\tilde{a}_2\rm,\nonumber\\
Q&\equiv&\left(\frac{\tilde{a}_3}{2}\right)^3-\left(\frac{\tilde{a}_3}{2}\right)\tilde{a}_2+\tilde{a}_1\rm,\nonumber\\\
R&\equiv&-3\left(\frac{\tilde{a}_3}{4}\right)^4+\left(\frac{\tilde{a}_3}{4}\right)^2\tilde{a}_2-\left(\frac{\tilde{a}_3}{4}\right)\tilde{a}_1+\tilde{a}_0\nonumber\rm.
\eer
These coefficients allow the definition of the \textit{cubic resolvent}:

\bea
u^3+2Pu^2+(P^2-4R)u-Q^2 = 0\rm,
\label{cubres}
\eea
upon whose solutions the solutions of the original quartic (Eqn.~\ref{qrt}) depend. The product of the solutions of the cubic resolvent is $u_1u_2u_3 = Q^2$ (Vieta's theorem), which clearly must be positive. The characteristics of the quartic's roots depend on the nature of the roots of the cubic resolvent (see Tab.~\ref{charqsolns}).

\linespread{1}
\begin{table}
\caption[The character of the solutions to a quartic equation]{The dependence of the solutions to the
parent quartic on the solutions to the cubic resolvent.}
\begin{center}
\begin{tabular}{cc}
\hline\hline\relax\\[-1.7ex]
Solutions of the cubic resolvent & Solutions of the quartic equation\\\relax\\[-1.7ex]
\hline\relax\\[-1.7ex]
all real and positive & all real\\
all real, one positive & two complex conjugate (cc) pairs\\
one real, one cc pair & two real, one cc pair\\\relax\\[-1.7ex]
\hline
\label{charqsolns}
\end{tabular}
\end{center}
\end{table}

Given the solutions of the cubic resolvent $u_1$, $u_2$, and $u_3$, the solutions of the quartic (Eqn.~\ref{qrt}) are

\ber
x_1&=&\frac{1}{2}(\sqrt{u_1}+\sqrt{u_2}+\sqrt{u_3})-\frac{a_3}{4a_4}\rm,\nonumber\\
x_2&=&\frac{1}{2}(\sqrt{u_1}-\sqrt{u_2}-\sqrt{u_3})-\frac{a_3}{4a_4}\rm,\nonumber\\
x_3&=&\frac{1}{2}(-\sqrt{u_1}+\sqrt{u_2}-\sqrt{u_3})-\frac{a_3}{4a_4}\rm,\nonumber\\
x_4&=&\frac{1}{2}(-\sqrt{u_1}-\sqrt{u_2}+\sqrt{u_3})-\frac{a_3}{4a_4}\rm.
\label{qsol}
\eer

\subsection{Solving a cubic equation}
The equations of the previous section reduce the problem of solving a quartic equation to that of solving a cubic equation (i.e. the cubic resolvent of Eqn.~\ref{cubres}).

Once again following \cite{bro97} (note the similarity to the method in the previous section), given a cubic equation:

\bea
b_3u^3+b_2u^2+b_1u+b_0 = 0,\;b_n\in\Re,\;b_3\neq0\rm,
\label{cub}
\eea
normalizing the equation and making the substitution $v = u + b_2/3b_3$ results in the reduced form:

\ber
v^3+pv+q & = & 0\rm,\nonumber
\eer
where, defining $\tilde{b}_n \equiv b_n/b_3$:

\ber
p&\equiv&-\frac{1}{3}\tilde{b}_2^2+\tilde{b}_1\rm,\nonumber\\
q&\equiv&2\left(\frac{\tilde{b}_2}{3}\right)^3-\left(\frac{\tilde{b}_2}{3}\right)\tilde{b}_1+\tilde{b}_0\rm.\nonumber
\eer
These coefficients allow the definition of the \textit{discriminant}:

\ber
D &\equiv &\left(\frac{p}{3}\right)^3+\left(\frac{q}{2}\right)^2\rm,
\label{disc}
\eer
upon which the characteristics of the solutions of the cubic equation depend (see Tab.~\ref{charcsolns}).

\linespread{1}
\begin{table}
\caption[The character of the solutions to a cubic equation]{The dependence of the solutions of a cubic equation on the
sign of the discriminant (assuming a real variable).}
\begin{center}
\begin{tabular}{cc}
\hline\hline\relax\\[-1.7ex]
D & Solutions of the cubic equation\\\relax\\[-1.7ex]
\hline\relax\\[-1.7ex]
$positive$ & one real, one complex conjugate pair\\
$negative$ & all real and distinct\\
$=0$ & all real, two (one, if $p=q=0$) distinct\\\relax\\[-1.7ex]
\hline
\label{charcsolns}
\end{tabular}
\end{center}
\end{table}

Given $p$, $q$, and $D$, Cardando's formula for the reduced form of the cubic leads to the solutions of the original cubic (Eqn.~\ref{cub}):

\ber
u_1&=&s+t-\frac{b_2}{3b_3}\rm,\nonumber\\
u_2&=&-\frac{1}{2}(s+t)-\frac{b_2}{3b_3}+i\frac{\sqrt{3}}{2}(s-t)\rm,\nonumber\\
u_3&=&-\frac{1}{2}(s+t)-\frac{b_2}{3b_3}-i\frac{\sqrt{3}}{2}(s-t)\rm,
\label{csolsc}
\eer
where:

\ber
s&\equiv&\sqrt[3]{-\frac{1}{2}q+\sqrt{D}}\rm,\nonumber\\
t&\equiv&\sqrt[3]{-\frac{1}{2}q-\sqrt{D}}\rm,\nonumber\\
i&\equiv&\sqrt{-1}\rm.\nonumber
\eer

If $D\leq 0$, the the cubic has three real roots, subject to the following two subcases, and the four real roots of the quartic follow directly from Eqn.~\ref{qsol}. If $D=0$, then $s=t$ and the cubic has three real solutions that follow directly from Eqn.~\ref{csolsc} from which one can see that two are degenerate. If $D<0$, the cubic has three distinct real roots. Obtaining these solutions via Eqn.~\ref{csolsc} requires intermediate complex
arithmetic. However, this may be circumvented by making the substitutions:

\ber
r&=&\sqrt{-\left(\frac{p}{3}\right)^3}\nonumber\\
\cos(\phi)&=&-\frac{q}{2r}\rm,\nonumber
\eer
in which case the solutions of the cubic (Eqn.~\ref{cub}) are:
\ber
u_1&=&2\sqrt[3]{r}\cos\left(\frac{\phi}{3}\right)-\frac{b_2}{3b_3}\rm,\nonumber\\
u_2&=&2\sqrt[3]{r}\cos\left(\frac{\phi+2\pi}{3}\right)-\frac{b_2}{3b_3}\rm,\nonumber\\
u_3&=&2\sqrt[3]{r}\cos\left(\frac{\phi+4\pi}{3}\right)-\frac{b_2}{3b_3}\rm.
\label{csolssub}
\eer
If $D>0$, then the cubic has one real root and a pair of complex conjugate roots and the quartic has two real roots and a pair of complex conjugate roots (see Tab.~\ref{charqsolns}). Finding the roots of the quartic involves intermediate complex arithmetic which may be circumvented as follows. Defining:

\ber
R&\equiv&-\frac{1}{2}(s+t)-\frac{b_2}{3b_3}\rm,\nonumber\\
C&\equiv&\frac{\sqrt{3}}{2}(s-t)\rm,\nonumber
\eer
Eqn.~\ref{csolsc} may be rewritten as:

\ber
u_1&=&s+t-\frac{b_2}{3b_3}\rm,\nonumber\\
u_2&=&R+iC\rm,\nonumber\\
u_3&=&R-iC\rm.\nonumber
\eer
Next, we have $u_{2,3} = \sqrt{R^2+C^2}e^{\pm iC/R}$. We then obtain the roots of the quartic from Eqn.~\ref{qsol}:

\ber
x_{1,2}&=&\frac{\sqrt{u_1}}{2}-\frac{a_3}{4a_4}\pm\sqrt[4]{R^2+C^2}\cos\left({\frac{C}{2R}}\right)\rm,\nonumber\\
x_{3,4}&=&\frac{-\sqrt{u_1}}{2}-\frac{a_3}{4a_4}\pm i\sqrt[4]{R^2+C^2}\sin\left({\frac{C}{2R}}\right)\rm.
\label{qsolfin}
\eer
Note that $x_1$ and $x_2$ are the two real solutions.

\section{The refined primitives algorithm}\label{solver}
Using the method above we created a SRHD primitive algorithm called ``REST\_ FRAME''. Given the speed advantage of the iterative root finder (see $\S$\ref{timing}), it a desirable choice over the analytical method within its regime of applicability, i.e. for low Lorentz factors. As Fig.~\ref{nr_acc} shows, the iterative root finder is accurate to order 10$^{-4}$ (see $\S$\ref{solacc}) for a sizable region of parameter space including all $R/E$ above the diagonal line between the points (0, $-7$) \& (9, 0) in the $\log(R/E)$ vs. $-\log(1-M/E)$ plane (i.e. for $\log(R/E) \ge -(7/9)\times\log(1-M/E)-7$). Therefore, for a given $M/E$ and $R/E$, we check if this inequality is true; if (not) so, we call the (analytical) iterative root finder (see $\S$\ref{pcode}).

\subsection{Pseudo-code}\label{pcode}
REST\_FRAME calculates the primitive variables given the conservative variables and the adiabatic index as represented in the following pseudo-code (note this is a 2D example):

\hspace*{8mm}PROCEDURE REST\_FRAME\\
\hspace*{8mm}RECEIVED FROM PARENT PROGRAM: $Y$, $Z$\\
\hspace*{8mm}RETURNED TO PARENT PROGRAM: $\gamma$, $v$, $C$\\
Comment: recall $Y\equiv M/E$ and $Z\equiv R/E$\\
Comment: $C$ is returned $< 0$ for code failures\\
\hspace*{8mm}GLOBAL VARIABLE: $\Gamma$\\
\hspace*{8mm}SET VALUE OF $m_{underflow}$\\
\hspace*{8mm}SET VALUE OF $v_{tol}$\\
Comment: determines iterative method velocity accuracy\\
Comment: we set $v_{tol}$ = 10$^{-8}$, 10$^{-10}$, 10$^{-12}$, 10$^{-14}$\\
Comment: for $-\log(1-Y) < 8.3$, $< 10.3$, $< 12.3$, otherwise, respectively\\
\hspace*{8mm}SET $M$ = $\sqrt{M_x^2+M_y^2}$\\
\hspace*{8mm}IF $M < m_{underflow}$ THEN\\
\hspace*{8mm}\hspace*{8mm}$v$ = 0, $\gamma$ = 1\\
Comment: avoids code failure if $v$ is numerically zero\\
\hspace*{8mm}ELSE\\
\hspace*{8mm}\hspace*{8mm}TEST FOR UNPHYSICAL PARAMETERS\\
\hspace*{8mm}\hspace*{8mm}IF PASSED, SET $C$ NEGATIVE AND RETURN\\
\hspace*{8mm}\hspace*{8mm}IF $\log(Z) \ge -(7/9)\times\log(1-Y)-7$, THEN\\
Comment: check to see if input parameters are within the acceptable\\ 
Comment: accuracy region of the iterative routine\\
\hspace*{8mm}\hspace*{8mm}\hspace*{8mm}CALL ITERATIVE\_QUARTIC($Y,Z,v_{tol},v$,$C$)\\
Comment: updates $v_{n-1}$ to $v_{n}$ using $n$ cycles of Newton-Raphson iteration\\
Comment: returns $v$ = $v_{n}$ when $|v_{n}-v_{n-1}|$ $\le$ $v_{tol}$\\
\hspace*{8mm}\hspace*{8mm}\hspace*{8mm}IF $C < 0$, THEN\\
Comment: this means the iteration failed to converge\\
\hspace*{8mm}\hspace*{8mm}\hspace*{8mm}\hspace*{8mm}RETURN\\
\hspace*{8mm}\hspace*{8mm}\hspace*{8mm}ELSE \\
\hspace*{8mm}\hspace*{8mm}\hspace*{8mm}\hspace*{8mm}$\gamma$ = $\sqrt{\frac{1}{1-v^2}}$\\
\hspace*{8mm}\hspace*{8mm}\hspace*{8mm}END IF\\
\hspace*{8mm}\hspace*{8mm}ELSE\\
\hspace*{8mm}\hspace*{8mm}\hspace*{8mm}CALL ANALYTICAL\_QUARTIC($Y,Z,\gamma$)\\
Comment: calculates $\gamma$ using analytical solution -- see below\\
\hspace*{8mm}\hspace*{8mm}\hspace*{8mm}$v$ = $\sqrt{1-\frac{1}{\gamma^2}}$\\
\hspace*{8mm}\hspace*{8mm}END IF\\
\hspace*{8mm}END IF\\
\hspace*{8mm}END PROCEDURE REST\_FRAME\\

\hspace*{8mm}PROCEDURE ANALYTICAL\_QUARTIC\\
Comment: see $\S$\ref{solveqe} for equations\\
\hspace*{8mm}RECEIVED FROM PARENT PROGRAM: $Y,Z$\\
\hspace*{8mm}RETURNED TO PARENT PROGRAM: $\gamma$\\
\hspace*{8mm}GLOBAL VARIABLE: $\Gamma$\\
\hspace*{8mm}$\tilde{a}_3$ = $2\Gamma(\Gamma-1)Z(Y^{-2}+1)$\\
\hspace*{8mm}$\tilde{a}_2$ = $(\Gamma^2-2\Gamma(\Gamma-1)Y^2-(\Gamma-1)^2Z^2)(Y^{-2}+1)$\\
\hspace*{8mm}$\tilde{a}_1$ = $-a_3$\\
\hspace*{8mm}$\tilde{a}_0$ = $(\Gamma-1)^2(Y^2+Z^2)(Y^{-2}+1)$\\
\hspace*{8mm}$\tilde{a}_4$ = $1+Y^2-a_0-a_2$\\
Comment: coefficients recast to counter subtractive cancellation -- see $\S$\ref{solacc}\\
\hspace*{8mm}NORMALIZE COEFFICIENTS TO $a_4$\\
Comment: e.g., $a_{3N}$ = $a_3/a_4$\\
\hspace*{8mm}CALCULATE CUBIC RESOLVENT COEFFICIENTS \\
\hspace*{8mm}CALCULATE DISCRIMINANT, $D$\\
\hspace*{8mm}IF $D\le$0 THEN\\
\hspace*{8mm}\hspace*{8mm}WRITE ERROR MESSAGE AND STOP\\
Comment: exploration suggests $D\le 0$ is unphysical but formal proof is elusive\\
Comment: thus, we leave $D\le 0$ uncoded with a error flag just in case\\
\hspace*{8mm}ELSE\\
Comment: $D>0$ $\Rightarrow$ $Q(\gamma)$ has 2 real roots (see Tab.~\ref{charqsolns} \& \ref{charcsolns})\\
\hspace*{8mm}\hspace*{8mm} CALCULATE ROOTS OF CUBIC RESOLVENT\\
Comment: the cubic has one real root and a pair of complex conjugate roots\\
\hspace*{8mm}\hspace*{8mm} IF REAL ROOT $< 0$, SET REAL ROOT = 0\\
Comment: the real root cannot be less than zero analytically\\
Comment: numerically, however, it can have a very small negative value\\
\hspace*{8mm}\hspace*{8mm} CALCULATE THE TWO REAL ROOTS OF THE QUARTIC\\
\hspace*{8mm}\hspace*{8mm} TEST FOR TWO OR NO PHYSICAL ROOTS\\
\hspace*{8mm}\hspace*{8mm} IF PASSED, WRITE ERROR MESSAGE, AND RETURN\\
\hspace*{8mm}\hspace*{8mm} IF FAILED, SET $\gamma$ = PHYSICAL ROOT\\
\hspace*{8mm}END IF\\
\hspace*{8mm}END PROCEDURE ANALYTICAL\_QUARTIC\\

\subsection{Code timing}\label{timing}
Using the Intel Fortran library function CPU\_TIME, we calculated the
CPU time required to execute 5$\times$10$^7$ calls to REST\_FRAME for
$Y$ = 0.9975 \& $Z$ = 1$\times 10^{-4}$ ($\gamma \sim$10) using the
Newton-Raphson iterative method with $Q(v)$ and 8-byte arithmetic,
and the analytical method with $Q(\gamma)$ and both 8-byte \& 16-byte
arithmetic (we investigated the use of 16-byte arithmetic due to an
issue with subtractive cancellation -- see $\S$\ref{solacc}). The CPU
time for each of these scenarios was 29.5, 36.5 (averaged over ten runs
and rounded to the nearest half second), and $\sim$11650 seconds
(one run only), respectively. This indicates that while using the 8-byte
analytical method is satisfactory, it is advantageous to use the
iterative method when Lorentz factors are sufficiently low, and that
the use of 16-byte arithmetic is a nonviable option. This result is
not surprising as the accuracy of Newton-Raphson iteration improves
by approximately one decimal place per iterative step \citep{dun94}
and the relative inefficiency of 16-byte arithmetic is a known issue \citep[e.g.][]{per06}.

\subsection{Solver Accuracy}\label{solacc}
The input parameters for our
primitives algorithm are the ratios of the laboratory-frame momentum
and mass densities to the laboratory-frame energy density (recall $Y
\equiv M/E$ and $Z \equiv R/E$) both of which must be less than
unity in order for solutions of Eqn.~\ref{lt} to exist. In addition,
the condition $Y^2 + Z^2 < 1$ must be met. Along with the fact that
$Y$ and $Z$ must also be positive, this defines the comprehensive
and physical input parameter space to be $0 < Y,Z < 1$ such that
$Y^2 + Z^2 < 1$ (we identify a particular region of parameter space
applicable to pulsar winds in the next section). We tested the accuracy of our 
iterative and hybrid primitives algorithms within this space as follows.

First, as we are most interested in light, highly relativistic flows 
(i.e. $Z$ small and $Y$ close to unity), to define the accuracy-search space 
we elected to use the quantities $-\log(1-Y)$, which for values greater 
than unity gives $0.9 < Y < 1$, and  $\log(Z)$, which for values less than 
negative unity gives $Z \ll 1$. We selected $0 <
-\log(1-Y) < 13$ and $-13 < \log(Z) < 0$ corresponding to Lorentz
factors ($\gamma$) between 1 and $2\times10^6$. We chose a range
with a maximal $\gamma$ slightly above $1\times10^6$ in order to
completely bound the pulsar wind nebula parameter space defined in the next
section.

Choosing a relativistic equation of state $\Gamma$= 4/3 and using 1300 points
for both $-\log(1-Y)$ and $\log(Z)$, we tested the accuracy of
REST\_FRAME by passing it $Y$ and $Z$, choosing $E = 1$, and using
the returned primitive quantities to derive the calculated energy
density $Ec$, and calculating the difference $|1-Ec/E| \equiv \delta
E/E$. We chose this estimate of the error because $\delta E/E \sim
\delta\gamma/\gamma$ and $\delta\gamma/\gamma$ is tied to the
accuracy of the numerical, hydrodynamic technique (see the final
paragraph in this section).

\begin{figure}[hb]
\caption{The accuracy (estimated as $\delta E/E$) of the
Newton-Raphson (N-R) iterative primitives algorithm where white,
light grey, medium grey, dark grey, and hatched regions correspond,
respectively, to an accuracy of order at least 10$^{-4}$, at least
10$^{-3}$, worse than 10$^{-3}$, failure, and unphysical input
($R^2/E^2 \ge 1-M^2/E^2$). Note that the Lorentz factor varies from
order 1 at the far left to order 10$^6$ at the far right. There is a
sizable white region representing $M/E < 0.999999$ ($\gamma < 500$)
and $R/E > 5\times 10^{-8}$ within which accuracy is generally
significantly better than 10$^{-4}$. N-R iteration is unreliable due
to sporadic failures for all $M/E$ and $R/E$ such that $R/E <
5\times 10^{-8}$ and for an ever increasing fraction of $R/E >
5\times 10^{-8}$ as $M/E$ increases until accuracy becomes
unacceptable or the code fails outright for $M/E$ and $R/E$ such
that $M/E > 0.999999$. Failures are due to divide by zero (see
\S\ref{sfix}) or nonconvergence within a reasonable number of
iterations.} \centerline{\includegraphics[width=4in]{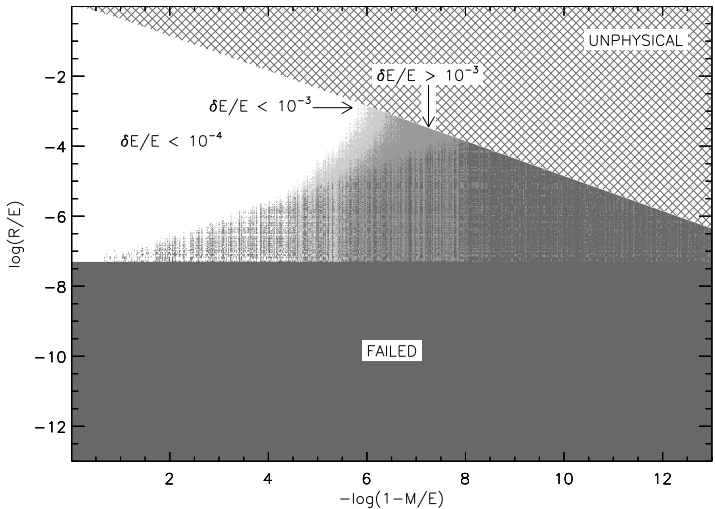}}
\label{nr_acc}
\end{figure}
\begin{figure}[hb]
\caption{The accuracy (estimated as $\delta E/E$) of the hybrid
primitives algorithm where white, light grey, and hatched regions
correspond, respectively, to an accuracy of order at least
10$^{-4}$, at least 10$^{-3}$, and nonphysical input ($R^2/E^2 \ge
1-M^2/E^2$). Note that the Lorentz factor varies from order 1 at the
far left to order 10$^6$ at the far right. The space between the
parallel lines represents PWNe input parameter space. The accuracy
degradation at the extreme right is due to subtractive cancellation
in the 4$^{th}$-order coefficient of the Lorentz-factor quartic as
$M/E\rightarrow$1.}
\centerline{\includegraphics[width=4in]{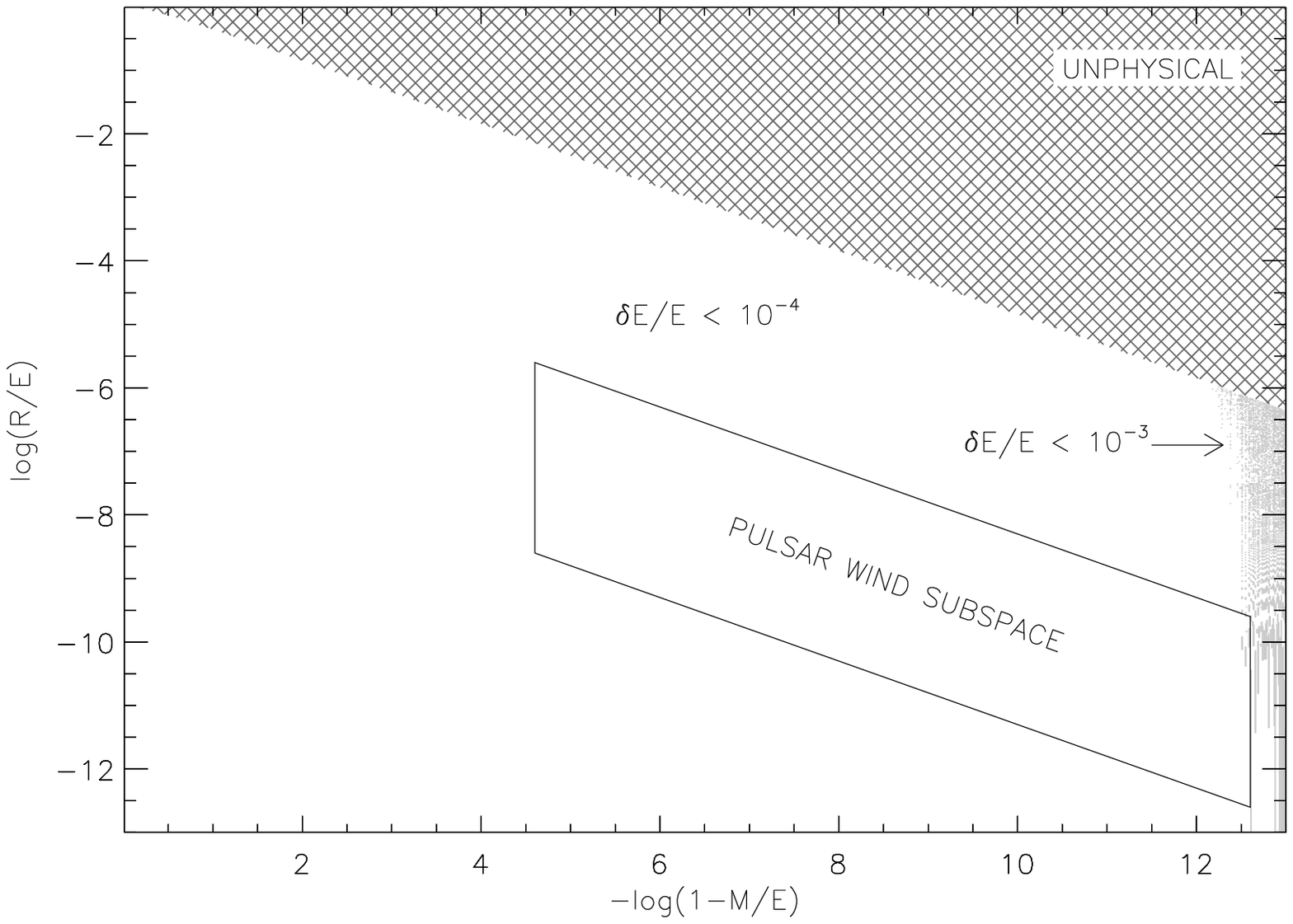}} \label{hy_acc}
\end{figure}

\begin{figure}[hb]
\caption{The accuracy (estimated as $\delta E/E$) of the
ultra-relativistic approximation of the flow velocity where white,
light grey, medium grey, and hatched regions correspond to an
accuracy of order at least 10$^{-4}$, at least 10$^{-3}$, worse than
10$^{-3}$, and unphysical input ($R^2/E^2 \ge 1-M^2/E^2$),
respectively. Note that the Lorentz factor varies from order 1 at
the far left to order 10$^6$ at the far right. The accuracy
degradation at the extreme right is due to the fact that the
fractional error in the Lorentz factor is proportional to the
fractional error in the velocity divided by $1-v^2$ which diverges
as $v\rightarrow$1.}
\centerline{\includegraphics[width=4in]{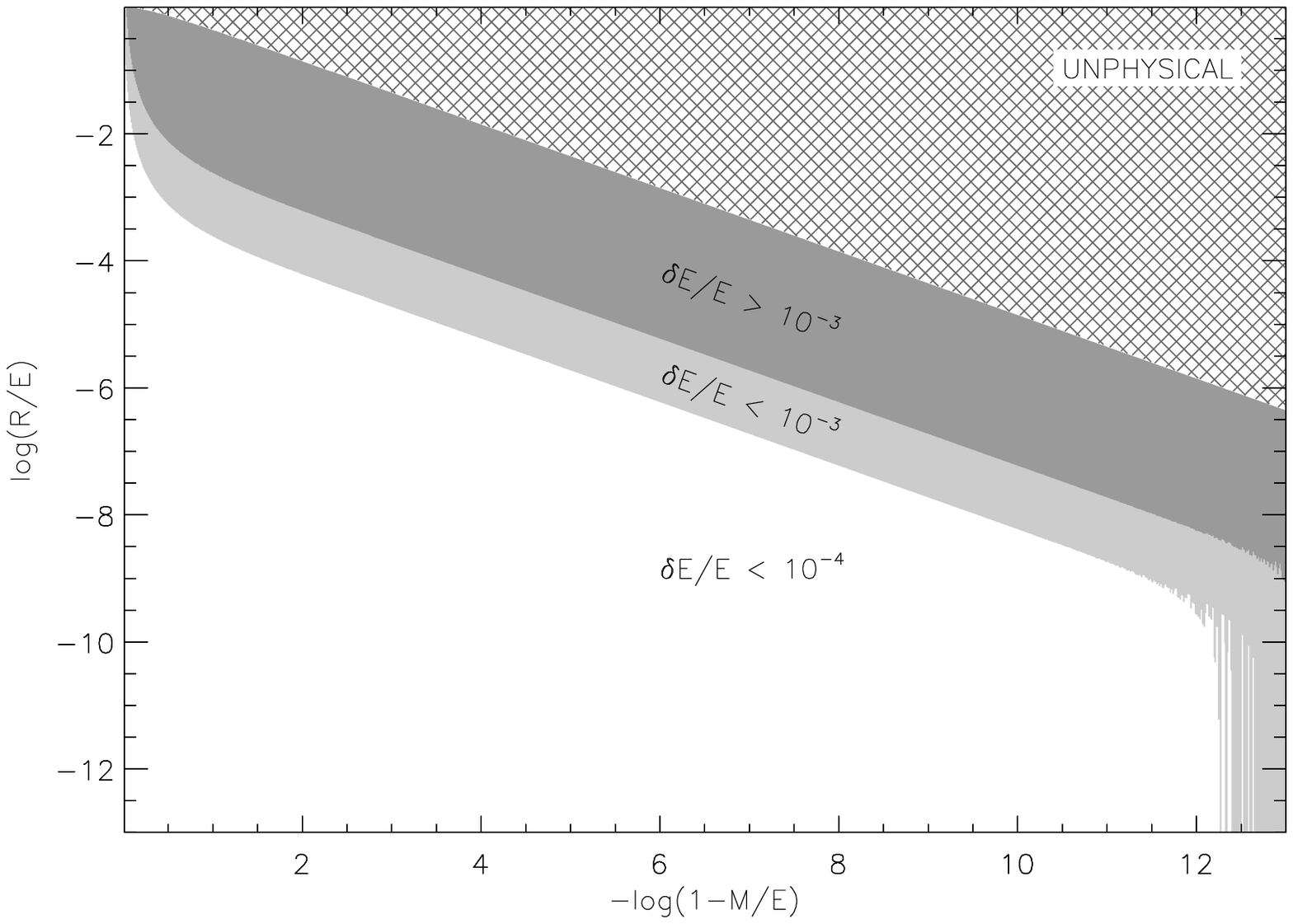}} \label{ur_acc}
\end{figure}

Our results for the Newton-Raphson (N-R) and hybrid methods are given in 
Figs.~\ref{nr_acc} \& \ref{hy_acc} which show where the accuracy is of order at 
least 10$^{-4}$, at least
10$^{-3}$, worse than 10$^{-3}$, failure, and unphysical input ($Z^2
\ge 1-Y^2$), respectively. We chose an accuracy of order 10$^{-4}$
as the upper cutoff because N-R iteration returns accuracies on this
order for $\gamma < 50$ and relativistic, hydrodynamic simulations
of galactic jets by \cite{dun94} produced robust
results for Lorentz factors of at least 50 using N-R iteration. An
additional result of interest is that the ultra-relativistic
approximation for $v$ (i.e. taking $R = 0$ thereby reducing $Q(v) =
0$ to a quadratic equation) manages an accuracy of at least
10$^{-4}$ for a large portion of the physical $Y-Z$ plane (see
Fig.~\ref{ur_acc}).

Fig.~\ref{nr_acc} shows the accuracy of the N-R iterative method.
There are several noteworthy features. First is the presence of a
sizable region corresponding to $\gamma < 500$ within which accuracy is generally significantly
better than 10$^{-4}$. Second is that N-R iteration is unreliable
due to sporadic failures for increasing Lorentz factors until accuracy becomes 
unacceptable or the code fails outright due to divide by zero (see \S\ref{sfix}) or
non-convergence within a reasonable number of iterations. In
addition, though N-R iteration has been widely established as the
primitives recovery method of choice for flows with Lorentz factors
less than order 10$^2$, we found that for a subset of parameters, corresponding to $\gamma < 2$, 
our N-R algorithm suffered an unacceptable degradation in
accuracy. The key to this problem lies in the how the 
flow velocity ($v$) is initially estimated for the first iterative cycle as 
follows:
\begin{enumerate}
 \item The established approach \citep{dun94,sch93} is to bracket $v$ with

\ber
v_{max} &=& min(1,Y+\delta)\rm,\nonumber\\
v_{min} &=& \frac{\Gamma -
\sqrt{\Gamma^2-4(\Gamma-1)Y^2}}{2Y(\Gamma-1)}\rm,
\eer
where $\delta \sim 10^{-6}$ and $v_{min}$ is derived by taking the ultra-relativistic limit (i.e. $R = 0$)
\item The initial velocity is then $v_i = (v_{min}+v_{max})/2+\eta$, where $\eta = (1-Z)(v_{min}-v_{max})$ for $v_{max} > \epsilon$ and $\eta = 0$ otherwise ($\epsilon$ order 10$^{-9}$)
\item This method fails due to selection of the incorrect root when the roots converge.
\item Thus, we make a simpler initial estimate of $v_i = v_{max}$, which guarantees that $v_i$ is ``uphill'' from $v$ for all physical $Y-Z$ space and that N-R iteration converges on $v$.
\end{enumerate}

Fig.~\ref{hy_acc} shows that our hybrid algorithm REST\_FRAME is
accurate to at least 10$^{-4}$ for all but a smattering of the
highest Lorentz factors. In fact, it is significantly more accurate over the
majority of the physical portion of the $Y-Z$ plane. The space
between the parallel lines represents the PWN input parameters
discussed in the next section. We find that multiplying
$Q(\gamma)$ by $(Y^2-Y^{-2})$ and rewriting the new $a_4$
($\tilde{a}_4$) in terms of the new $a_2$ ($\tilde{a}_2$) and new
$a_0$ ($\tilde{a}_0$), e.g. $\tilde{a}_4 =
1+Y^2-\tilde{a}_0-\tilde{a}_2$, improves the accuracy somewhat, but
does not entirely mitigate the problem. The issue of accuracy loss
at large Lorentz factors in 8-byte primitives algorithms is a known
issue \citep[e.g.][]{nob03} for which we know of no complete 8-byte solution.
Employing 16-byte arithmetic provides spectacular accuracy, but
introduces an unacceptable increase in run time (see
$\S$\ref{timing}).

The issue of what constitutes an acceptable error in the calculated
Lorentz factor is decided by the fact that a fractional error in
$\gamma$ translates to the same fractional error in $p$ and $n$
which are needed to calculate the wave speeds that form the basis of
the numerical, hydrodynamic technique, a
Godunov scheme \citep{god59} which approximates the solution to the
local Riemann problem by employing an estimate of the wave speeds.
We do not know a priori how accurate this estimate needs to be,
and so procede with 8-byte simulations of pulsar winds
with the expectation of using shock-tube tests \citep{tho86} to 
validate the accuracy of the
computation of well-defined flow structures as we approach the
highest Lorentz factors. It is also noteworthy that while $\gamma$ =
10$^6$ is the canonical bulk Lorentz factor for pulsar winds,
$\gamma$ = 10$^4$ and 10$^5$ are still in the ultra-relativistic
regime, and it may very well prove to be that these Lorentz factors
are high enough to elucidate the general ultra-relativistic,
hydrodynamic features of such a system. The hybrid algorithm
achieves accuracies of at least 10$^{-6}$ for $\gamma \sim 10^5$,
which is safely in the acceptable accuracy regime.

\section{Application to Bow-shock Pulsar Wind Nebulae}\label{appl}
At the end of a massive star's life, the collapse of its core to a 
compact object, i.e. a neutron star or black hole, drives a shockwave 
into its outer layers, thereby heating and ejecting them into the 
interstellar medium (ISM) in a supernova (SN) explosion. Subsequently, 
the shockwave overtakes the ejecta, expanding into the ISM, and forming 
a supernova remnant (SNR). Typically, a SN releases \s10$^{51}$ erg 
of mechanical energy that drives expansion of the SNR, sweeping up ISM 
material, heating it to X-ray temperatures and infusing it with fusion 
products beyond lithium.

In a subclass of SNRs, for progenitor masses between 10 and 25 solar masses \citep[e.g.][]{hag03}, 
the compact object formed in the SN explosion is a 
rapidly-spinning, highly-magnetized neutron star surrounded by a 
magnetosphere of charged particles. The combination of the rotation and 
the magnetic field gives rise to extremely powerful electric fields that 
accelerate charged particles to high velocities. The magnetic field interacts 
with the charged particles resulting in the spin-down of the 
neutron star, and the release of spin-down energy. A relatively small 
fraction of this energy is converted into beamed emission, manifest as 
an apparent pulse if the neutron star's rotation sweeps the beam across 
the Earth, leading to the designation ``pulsar''. The bulk of the spin-down 
energy is converted into a pulsar wind \citep{mic69} which is terminated at a strong shock, downstream of which the flow is indistinguishable from
being spherically symmetric \citep[e.g.][]{cha02}. 
The wind particles interact with the 
magnetic field causing them to emit synchrotron radiation, forming a pulsar 
wind nebula (PWN). The Crab Nebula, formed in the SN explosion of 1054 CE, is the canonical object of this type. The Crab exhibits pulsations from the radio, all the way up to X-rays, and is a prodigious source of $\gamma$-rays.

The wind in the immediate vicinity of the pulsar is a diffuse, 
relativistic gas unlikely to be directly observable. However, the 
classic structure of forward and reverse shocks separated by a contact 
surface \citep{wea77} arises from the wind interaction with the SNR or 
ISM. A probe of this interaction is provided by optical emission from 
the swept-up ambient ISM, thermal X-ray emission from the SNR and/or the 
shocked ISM, and X-ray synchrotron emission from the shocked wind. 
Furthermore, the high space velocity that is typical of pulsars \citep{cor98}
implies an asymmetric ram pressure on the pulsar wind from the denser 
ambient medium. The details of the morphology and of the distribution of 
the density, pressure, and velocity within the PWN depends upon the 
density, speed, momentum, and energy flux of the pulsar wind. Thus, 
comparison of PWN simulations with observational data can provide an 
unparalleled method for investigating pulsar winds and, hence, how the 
surrounding medium taps the rotational energy of the pulsar.

\cite{pac73} and \cite{rg74} pioneered the basic model of PWNe; a model further developed by
\cite{kc84a,kc84b} and \cite{emm87}. \cite{gae06} and \cite{buc08} are excellent reviews on observational
and theoretical studies of PWNe, respectively. For a number of reasons, a detailed, 
quantitative study of PWNe is now 
particularly timely. First, there is a cornucopia of high-quality data 
from space-born observatories such as the $Chandra$ X-ray Observatory and 
XMM-Newton. Second, the total energy radiated by PWNe accounts for only a small fraction of the spin-down energy, leaving 
a large energy reservoir available for interaction with the SNR and 
acceleration of ions, the partitioning of which is not well understood.

Efforts to model PWNe span three decades \citep[with seminal papers][]{rg74,kc84a,kc84b,emm87}. While the case for a non-isotropic pulsar wind energy flux has long been made \citep{mic73}, it has only been recently that a theoretical explanation of the mechanism behind the jet/torus structure interior to the termination shock has been put forward, and that the predictions of \cite{mic73} have been confirmed \citep[][]{buc08}. In particular, \cite{buc08} highlighted that a detailed description has been made possible by the increase in the efficiency and robustness of relativistic, numerical MHD codes stemming from the work of \cite{kom99}, \cite{zan03} and \cite{gam03}. Simulations by \cite{zan04} indicate that where jet formation in PWNe takes place is tied to where the magnetic field attains equipartition, at which point the magnetic filed can no longer be compressed. If this happens close to the termination shock, then, due to the mildly relativistic nature of the post-shock flow, hoop stresses can become efficient and most of the flow is diverted back toward the axis and collimated. The magnitude of the magnetization is key: if it is too small, the equipartition is reached outside the nebula, hoop stresses remain inefficient, and no collimation is produced.

Other modeling, for example, that presented herein, is concerned with the \textit{global} structure of PWNe. The enormous acceleration of the wind at the termination shock smears out asymmetries leaving an essentially spherically symmetric shocked flow to produce large scale PWN features. In particular, \cite{buc05} and \cite{vig07} are two recent examples of simulations addressing the structures that arise in bow-shock PWNe (see below). \cite{buc05} were the first to apply a fully-relativistic MHD code \citep{zan03,zan02}, and, for an axisymmetric geometry, obtained a relativistic backflow behind the pulsar, as predicted by \cite{wan93} for PSR1929+10. However, the wind Lorentz factor and pulsar velocity were 10 and 9000 km s$^{-1}$, respectively, which are far from the typical values of 10$^6$ and 500 km s$^{-1}$ (indeed, the Guitar pulsar, the fastest known, has a transverse velocity of $\sim$1700 km s$^{-1}$). In addition, the paper does not address the ``bubble'' in the Guitar (see Fig.~\ref{fig:guitar}). \cite{vig07} performed \textit{non}-relativistic, hydrodynamic simulations with a relaxation to cylindrical symmetry. The full 3-D FLASH code \citep{fry00} was employed and an anisotropic pulsar wind, cooling of the shocked ISM, ISM density gradients, and ISM walls were considered. While the authors employed a realistic pulsar velocity of 400 km s$^{-1}$, the non-relativistic nature of the simulations limited the Lorentz factor to order unity. In this section, we present fully-relativistic, axisymmetric, hydrodynamic simulations of bow-shock PWNe for a realistic pulsar velocity and wind Lorentz factor. In particular, we address the origin of relativistic backflows leading to a persistent nebular bubble.

\subsection{Bow-shock formation}\label{chap-intro:pwnevo}
The evolution of PWNe can be broken into four broad phases:
1) free-expansion, 2) SNR reverse shock interaction, 3) expansion inside a Sedov SNR, and 4) bow shock formation \citep[for a detailed discussion, see][and references therein]{gae06}. In this work, we investigate the last stage of evolution. The time it takes for the pulsar to cross the SNR was obtained by \cite{swa03}:
\ber
t_{cr} = 1.4\times 10^4\left(\frac{E_0}{10^{51}\;\rm{erg}}\right)^{1/3}\left(\frac{v_{psr}}{10^3\;\rm{km s^{-1}}}\right)^{-5/3}\left(\frac{n_0}{1 \;\rm{cm}^{-3}}\right)^{-1/3}\rm,
\eer
where $v_{psr}$ is the velocity of the pulsar. Once the PWN-SNR system has evolved to the Sedov-Taylor stage, the time elapsed is sufficiently large that is possible for the pulsar to have reached the edge of the nebula, or even beyond \citep{swa01}. Thus, the pulsar escapes its original wind bubble, leaving behind a ``relic'' PWN, and traverses the SNR while inflating a new PWN. As the pulsar moves away from the center of the remnant, the sound speed decreases. Following \cite{swa98}, \cite{swa04} calculated the Mach number of the pulsar, $\mathcal{M}_{psr}$, and found that $\mathcal{M}_{psr}$ exceeds unity after a time $t = 0.5t_{cr}$, at which point the pulsar has travelled a distance $R_{psr} \simeq 0.677 R_{snr}$, and the nebula is deformed into a bow shock. The condition on the pulsar velocity for this transition to occur while the remnant is in the Sedov-Taylor phase is given by \citep[][and references therein]{swa04}:
\ber
v_{psr} \ge 325\left(\frac{E_0}{10^{51}\;\rm{erg}}\right)^{1/17}\left(\frac{n_0}{1 \;\rm{cm}^{-3}}\right)^{2/17}\rm{km\;s^{-1},}
\eer 
a relation showing a strikingly weak dependence on the physical parameters. A significant fraction (30--40\% depending on the velocity distribution model) of the pulsars compiled by \cite{arz02} satisfy this condition. \cite{swa03} showed that once the pulsar reaches the edge of the remnant, its Mach number is $\mathcal{M}_{psr} \simeq 3.1$. Subsequently, the pulsar moves through the ISM where its velocity corresponds to a hypersonic Mach number typically on the order of 10$^2$. 

\begin{figure}[ht]
\centerline{\includegraphics[angle=0,scale=0.6]{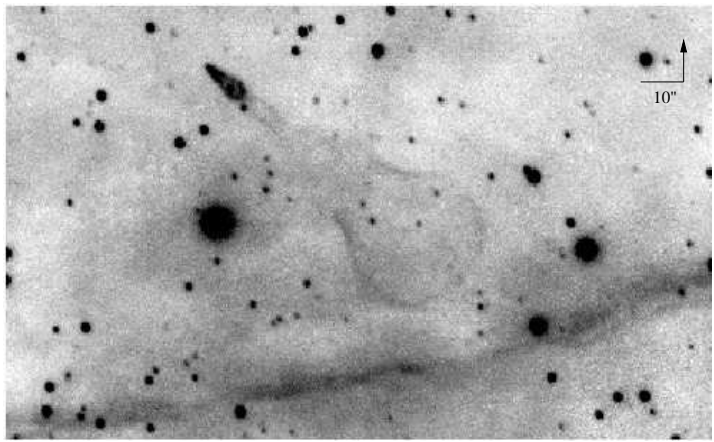}}
\caption{A 1995 Hale Telescope H$\alpha$ image of the Guitar Nebula (20 angstrom filter at 6564 angstroms). The cometary neck connecting to a spherical bubble are clearly evident. Credit: \cite{cha02}.}
\label{fig:guitar}
\end{figure}

\begin{figure}[ht]
\centerline{\includegraphics[angle=0,scale=0.75]{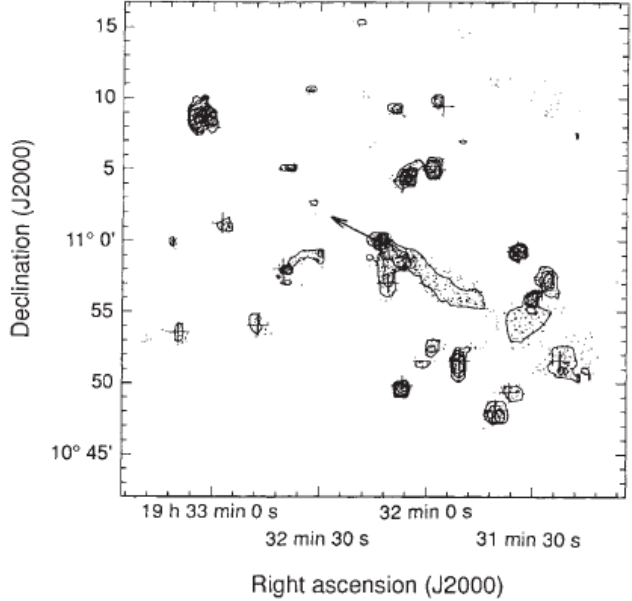}}
\caption{\textit{R}OSAT X-ray surface brightness in the field of PSR1929+10 showing the X-ray tail. \cite{wan93} suggested that the X-ray morphology is due to the acceleration of particles behind the pulsar forming a relativistic backflow. North is up and East is left. Credit: \cite{wan93}.}
\label{fig:1929}
\end{figure}

The most famous example of a PWN in this stage of evolution is the Guitar nebula \citep[][see Fig.~\ref{fig:guitar}]{cor93}, so named because of its cometary neck connecting to a nearly spherical bubble. Numerous other examples are shown in \cite{kar08}. A case of particular import to this work is that of the X-ray emission associated with PSR1929+10 (see Fig.~\ref{fig:1929}). \cite{wan93} posited that the morphology is due to a relativistic backflow behind the pulsar, a suggestion that has gone unconfirmed for realistic wind Lorentz factors and pulsar velocities, and was a prime motivator for this project. The simulations in this section directly probe the morphology and interior structure of PWNe during this phase, motivate how the shape of the Guitar nebula persists, without resorting to tailored ISM geometry, and confirm the interpretation of \cite{wan93}.

\subsection{Identifying suitable input parameters}\label{sec:hpar}
The outflow streams relativistically into the ambient medium generating a
strong shock. We derive a value for the outflow pressure, $p_o$, from the assumption that the outflow
is interacting with the ambient medium requiring that the momentum flux be
comparable on either side of this shock; if the fluxes were not comparable, then either the ambient flow or
outflow would dominate and the problem would be uninteresting. The momentum flux
of the non-relativistic ambient medium and ultra-relativistic outflow are, respectively:
\ber
F_{M,a}&=&n_av_a^2+p_a\rm,\nonumber\\
F_{M,o}&=&\gamma_o^2(e_o+p_o)v_o^2 + p_o\nonumber\rm.
\eer
For an ultra-relativistic outflow, $p_o\gg n_o\Rightarrow e_o\rightarrow 3p_o$,
and $v_o \rightarrow 1$, and, for the ambient medium, $n_av_a^2\gg p_a$.
Applying these conditions, and noting that $\gamma_o^2p_o\gg p_o$, gives:
\ber
p_o &\sim& n_a\left(\frac{v_a}{2\gamma_o}\right)^2 \sim 10^{-19}\;for\;\gamma_o\;=\;10^6,\;n_a=1\rm.\nonumber
\label{eqn:mbal}
\eer
We are then free to pick any $n_o$ meeting the conditions of a light,
relativistic outflow, i.e. $n_a,\;p_o \gg n_o$. This condition is motivated
by the fact that the flow is \textit{very} fast ($\gamma$ $>$ 10$^4$). Well below the length scale 
of this study, the flow will be stabilized by the strong magnetic field, synchrotron cooling will be 
strong, and adiabatic losses due to expansion across the orders of magnitude in scale between the pulsar and the termination
shock will sap internal energy. This will conspire to effectively stop
energy from being converted into thermal motions. Under those conditions, the flow might be cold. 
However, on the scale of the termination shock 
the field is much weaker meaning far less
synchrotron cooling, instability is less inhibited, and interesting evolution will occur over fewer 
adiabatic-loss scale lengths. Indeed, the flow might
well be influenced by waves generated both upstream and downstream of the
shock(s). The scales relevant to this work correspond to the region of ``hot'' post-termination shock plasma \citep[e.g.][]{buc08} within a PWN. Therefore, a hot flow seems significantly more plausible than does a cold flow
and we select $n_o = 10^{-l}p_o$, $3 < l < 6$. This clearly satisfies
$p_o \gg n_o$ and one may verify it satisfies $n_a \gg n_o$ by
noting that the equation for $p_o$ implies $n_a \gg p_o$
since $\gamma_o^2 \gg v_a^2$ for the flows of interest here.

\subsection{A relativistic backflow}
\begin{figure*}%
\captionsetup[figure]{margin=10pt}%
\caption{An 871,200-iteration simulation of a light,
ultra-relativistic outflow interacting with a dense, slow ambient
flow. The input parameters are: $v_a$ = 0.00583 ($\sim$1750 km s$^{-1}$), $\mathcal{M}$ = 300, $n_a$ = 1, $\gamma_o$ = 10$^{5}$, $p_o$=
7$\times$10$^{-16}$, and $n_o$ = 10$^{-3}p_o$. The upper and lower panels show a linear color map of the rest-frame pressure and lab-frame mass density, respectively. Both have been
reflected along the symmetry axis. The outflow originates within the circular
region to the left of the evolving structure and the ambient flow
streams in along the left edge of the domain. The lines labelled ``1''
and ``2'' are 1-D data cuts (hereafter ``cut-h1''and ``cut-h2'', respectively) with flow parameters plotted in Fig.~\ref{fig:vars}.}%
\subfloat[Linear pressure map.\label{sim-a}]{\includegraphics[angle=-90,width=5in]{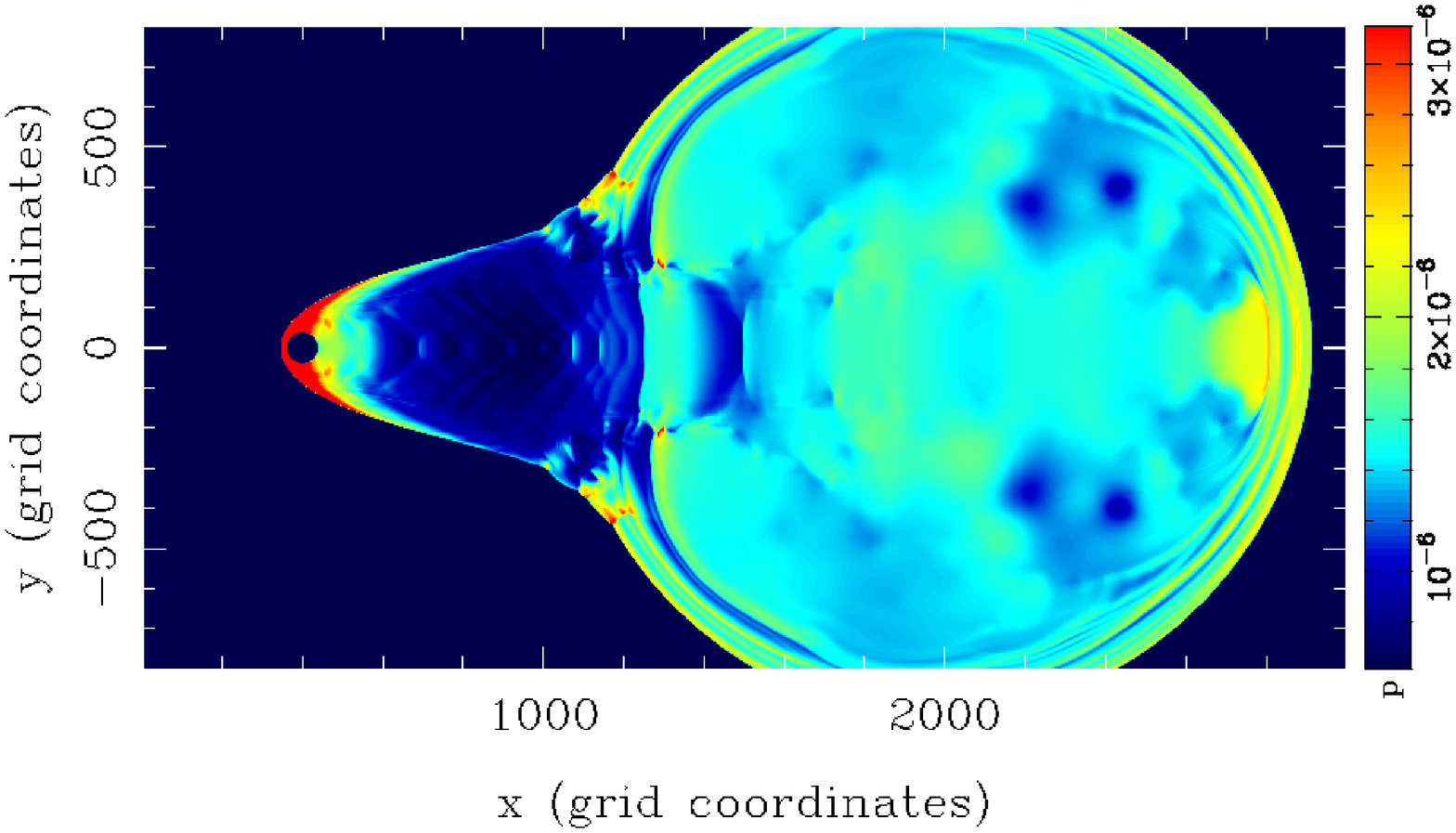}}\\
\hfill
\subfloat[Linear lab-frame mass density map.\label{sim-b}]{\includegraphics[angle=-90,width=5in]{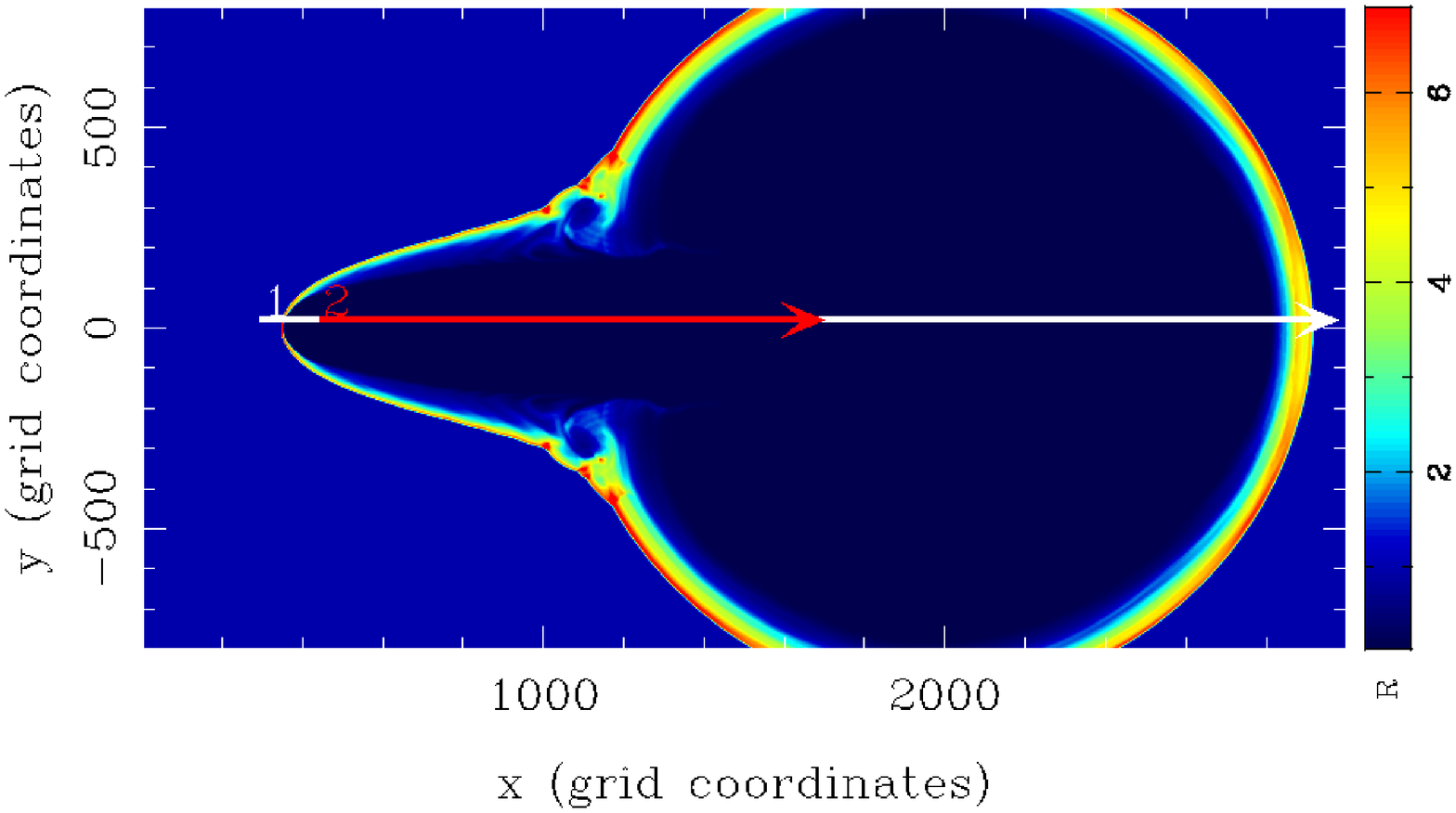}}\\[-10pt]
\label{sim}%
\end{figure*}

\begin{figure}[ht]
\caption{Plotted in red overlaying the pressure map for the simulation shown in Fig.~\ref{sim} is the refined grid at level L=1. The bottom half of the map is a reflection of the top half and has the same refined grid even though it is not shown. Note that the red lines trace the outlines of the meshes of refined cells, but not the cells themselves. While the boundary shock is well-refined, the axial shocks within the nebula are not refined at all. Flagging is determined by the largest difference in R between adjacent cells for all cells at level L. The refinement only follows the boundary shock because R differences inside the nebula are small compared to the difference between the nebula and the ambient medium. We will investigate refinement flagging in more detail in a future study.}
\vspace*{3mm}\centerline{\includegraphics[angle=-90,width=5in]{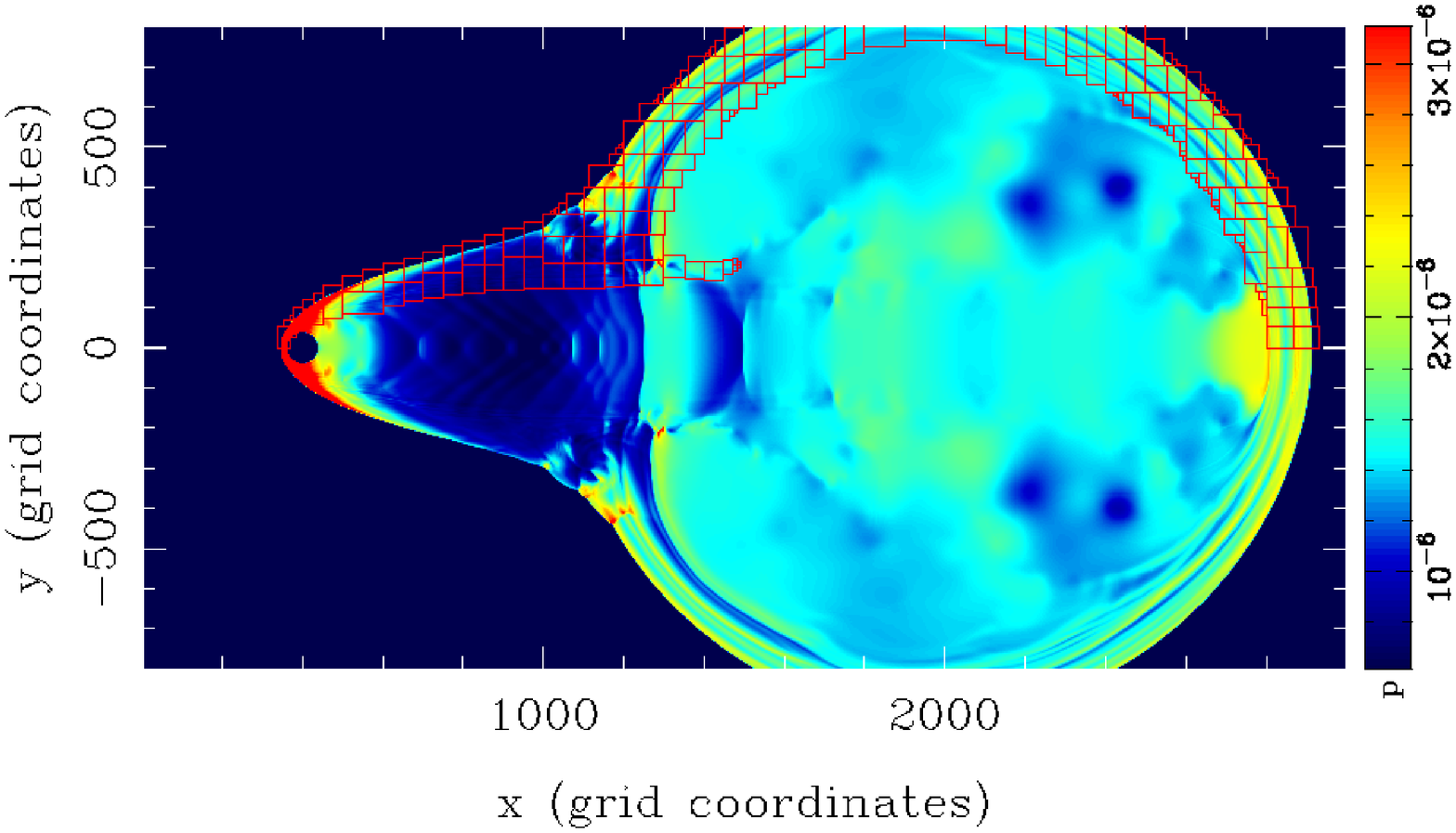}}
\label{fig:sim-adapt}
\end{figure}

Fig.~\ref{sim} shows a simulation of a $\gamma_o$ = 10$^{5}$ outflow
interacting with an ambient flow with velocity $v_a$ = 0.00583($\sim$1750 km~s$^{-1}$). The outflow
pressure was calculated for an ambient-flow velocity of 500 km~s$^{-1}$ in order
to match the typical value for pulsars in general.
The outflow originates inside the circular
region to the left of the evolving structure and the ambient flow streams in
along the left edge of the computational domain. Fig.~\ref{fig:sim-adapt} shows the limited extent 
of the refined grid, supporting the choice of a maximum number of refinement level of L\tsb{max} = 1. Recall the H$\alpha$ image of the 
Guitar Nebula (see Fig.~\ref{fig:guitar}), a well-known pulsar wind nebula with the most rapidly moving 
pulsar ever observed, with a transverse velocity of (1.7$\pm$0.4)$\times$10$^{3}$
km~s$^{-1}$ \citep{cha02}. The simulation qualitatively resembles the nebula. This result constitutes
compelling motivation for the conclusion that interstellar-medium flows set up by
the space motion of pulsars can indeed produce ``cometary'' nebulae.

\begin{figure}
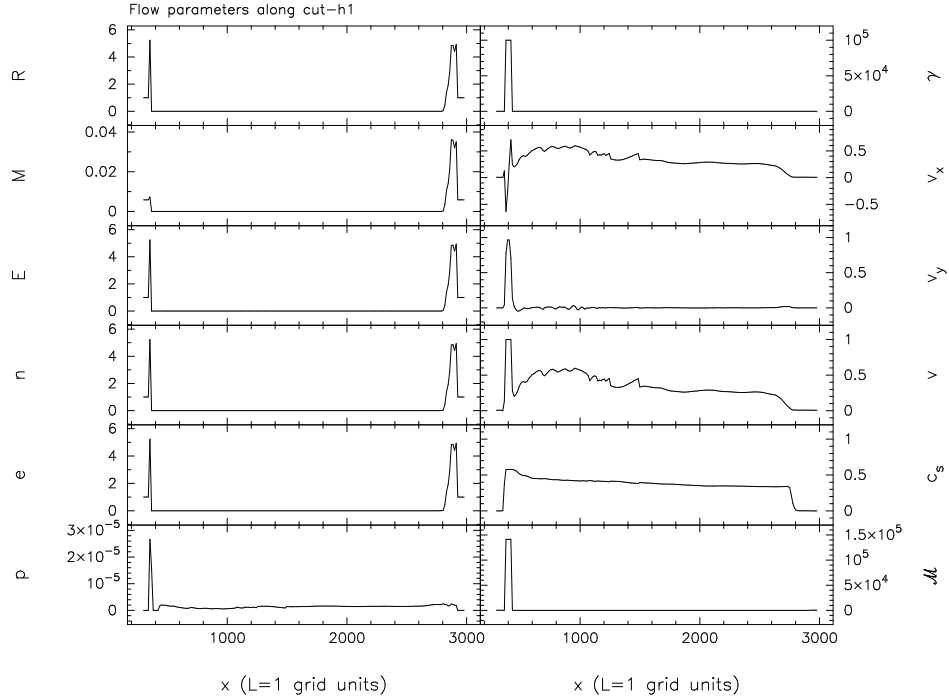

\captionsetup[figure]{margin=10pt}%
\caption{The run of the laboratory-frame mass (R), momentum (M), and total energy
(E) densities, rest-frame mass (n), and total energy (e) densities, and pressure (p), 
Lorentz factor ($\gamma)$, x- and y-components of the flow velocity (v$_{\rm x}$, v$_{\rm y}$), the flow velocity (v), sound
speed (c$_{\rm s}$), and generalized Mach number ($\mathcal{M}$) along (a) cut-h1 and (b) cut-h2 in Fig.~\ref{sim}.}
\subfloat[Variables along cut-h1.\label{fig:vars-a}]
{\includegraphics[angle=-90,width=4.9in]{fig9a.ps}}\\
\subfloat[Variables along cut-h2.\label{fig:vars-b}]
{\includegraphics[angle=-90,width=4.9in]{fig9b.ps}}\\
\label{fig:vars}
\end{figure}

We believe this simulation to be the first demonstrating
asymmetry arising from a spherically-symmetric, light,
ultra-relativistic flow interacting with a dense, slow ambient flow.
The lines labeled ``1'' and ``2'' on the density map in Fig.~\ref{sim} mark
one-dimensional cuts (hereafter ``cut-h1''and ``cut-h2'', respectively) made to probe the state of the simulation.
Cut-h1 spans the entire structure while cut-h2 spans
the interior space occupied by the pressure enhancements clearly
visible in the pressure map. Fig.~\ref{fig:vars} shows the
values of the flow parameters along these cuts. 
These plots clearly show the outer bounding shockwave
represented by the red boundary in the
density map as well as a series of weaker internal on-axis shocks visible
in the pressure map. The x-component of the flow velocity shows
that a relativistic back flow harboring a series of weak shocks has arisen down stream. This
validates the interpretation by \cite{wan93} of the origin of the X-ray trail behind PSR1929+10, and
demonstrates the ability of the refined solver to elucidate the internal
structure of diffuse, ultra-relativistic pulsar wind nebulae which is often difficult 
to observe directly.

It is noteworthy that the termination shock of the wind is not evident in the simulation discussed above. 
This is due to numerical shocking of the wind as it emerges from the on-axis hemisphere, as follows. 
Consider the cells depicted in Fig.~\ref{fig:ishock}. Let the angle of the line connecting the center of 
the hemisphere and the center of a cell 1, 2, 3, or 4 be $\theta_i$, $i$ = 1, 2, 3, 4. Since we have taken 
the pulsar wind to be spherically symmetric as it emerges from the hemesphere, we may calculate the relative 
flow velocities $\Delta v_{12}$ and $\Delta v_{34}$ (normalized to the speed of light) at the centers of cells 1 \& 2 and 3 \& 4:\newpage

\begin{figure}[ht]
 \caption{Schematic geometric cell layout pertaining to numerical shocking of the pulsar wind. The arc
 represents the on-axis hemisphere with radius 37.5 fine cells. Cell 1 is on-axis and is centered at 41.5 fine cells from the center of 
 the hemisphere (relative center coordinates (x,y) = (41.5,0.5). The center coordinates of cells 2, 3, and 4 are (41.5,1.5), (29.5,29.5), and (29.5, 30.5) respectively.}
 \vspace*{3mm}\centerline{\includegraphics[width=3.5in]{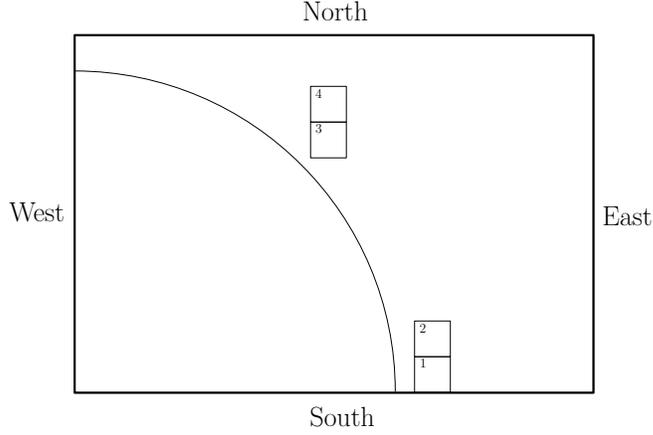}}
 \label{fig:ishock}
 \end{figure}

\begin{figure}[hb]
\captionsetup[figure]{margin=10pt}%
\caption{A time sequence of linear pressure maps for the simulation shown in Fig.~\ref{sim}. The sequence indicates that the appearance of the inflection point is preceded by a pressure drop inside the nebula. Note the finer time steps between 240K and 360K iterations and that the color map is relative to the minimum and maximum for each plot individually. However, the minimum is the same and the maximum is similar for all plots, so the variation is minimal.}%
\subfloat[10,000 iterations.\label{fig:evo-a}]
{\includegraphics[angle=-90,width=1.8in]{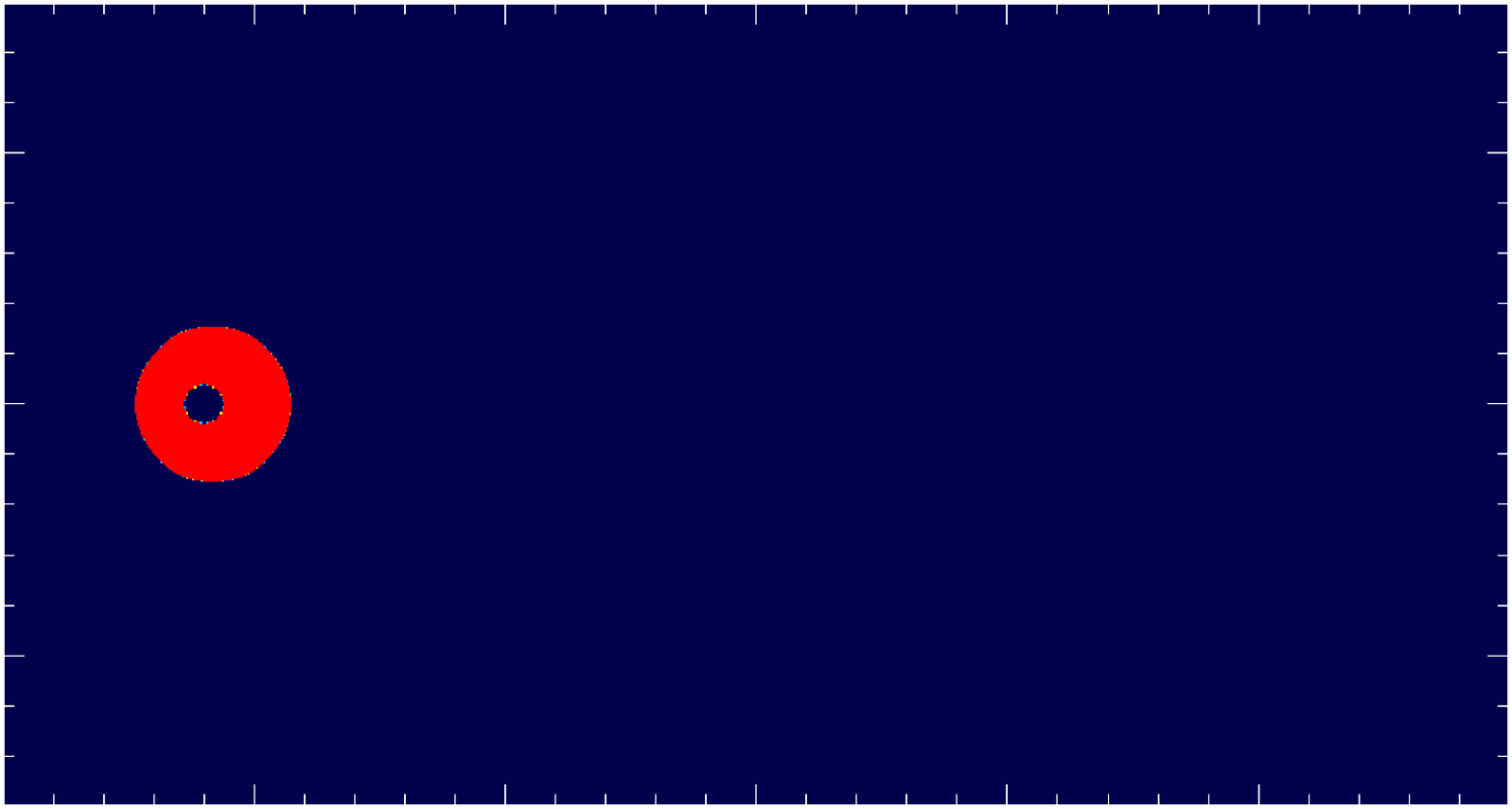}}
\subfloat[120,000 iterations.\label{fig:evo-b}]
{\includegraphics[angle=-90,width=1.8in]{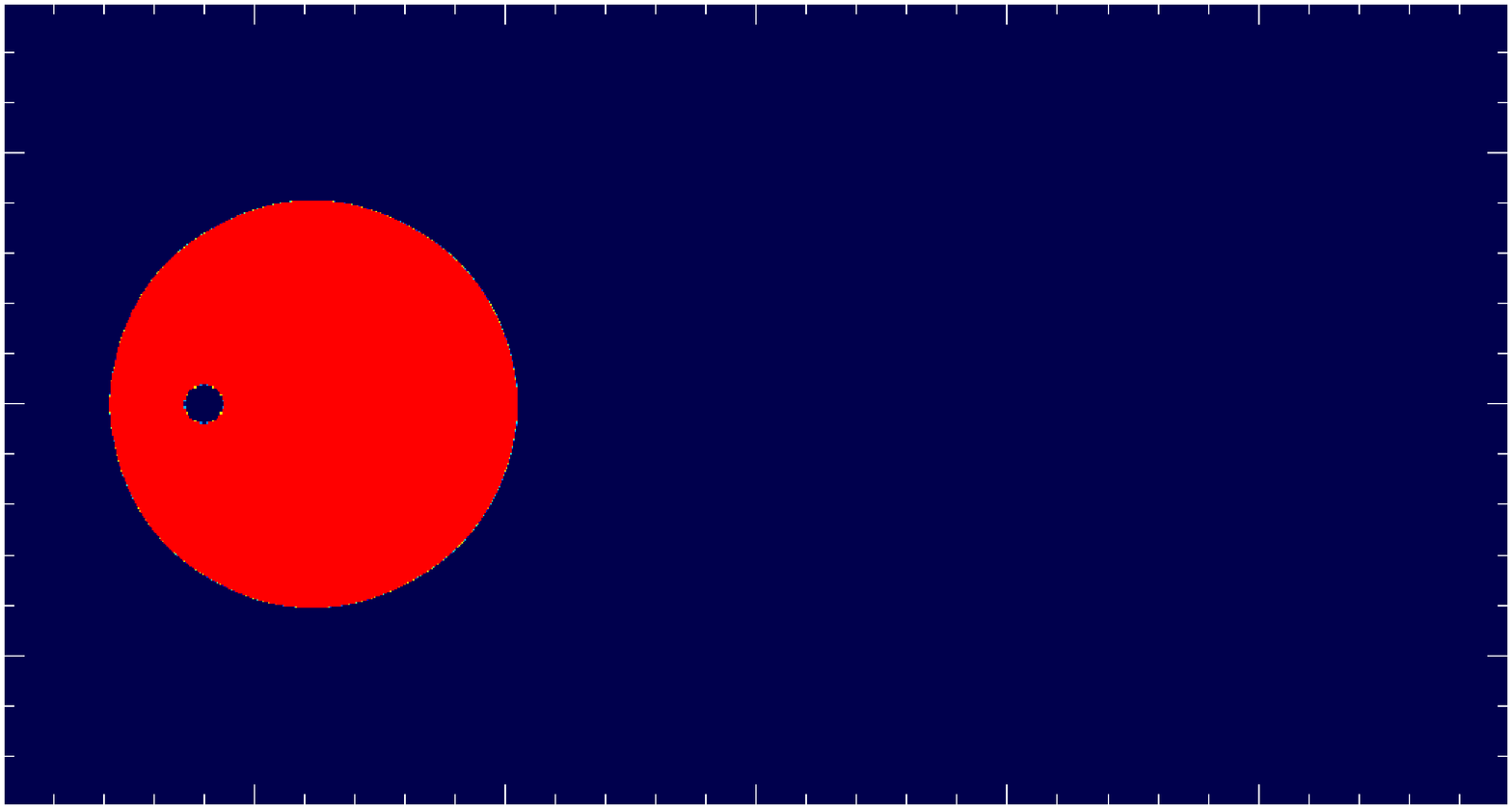}}
\subfloat[240,000 iterations.\label{fig:evo-c}]
{\includegraphics[angle=-90,width=1.8in]{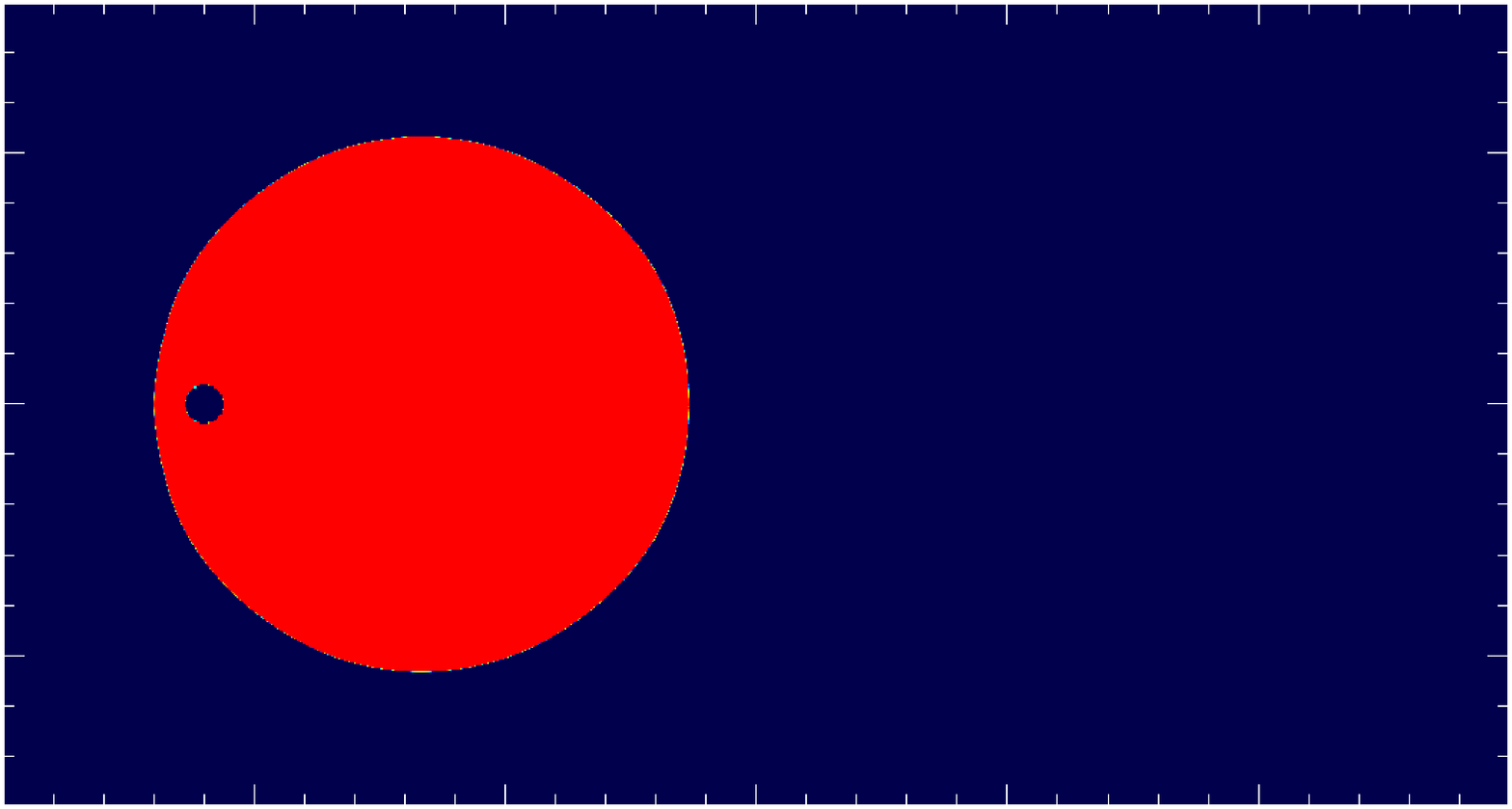}}\\
\subfloat[300,000 iterations.\label{fig:evo-d}]
{\includegraphics[angle=-90,width=1.8in]{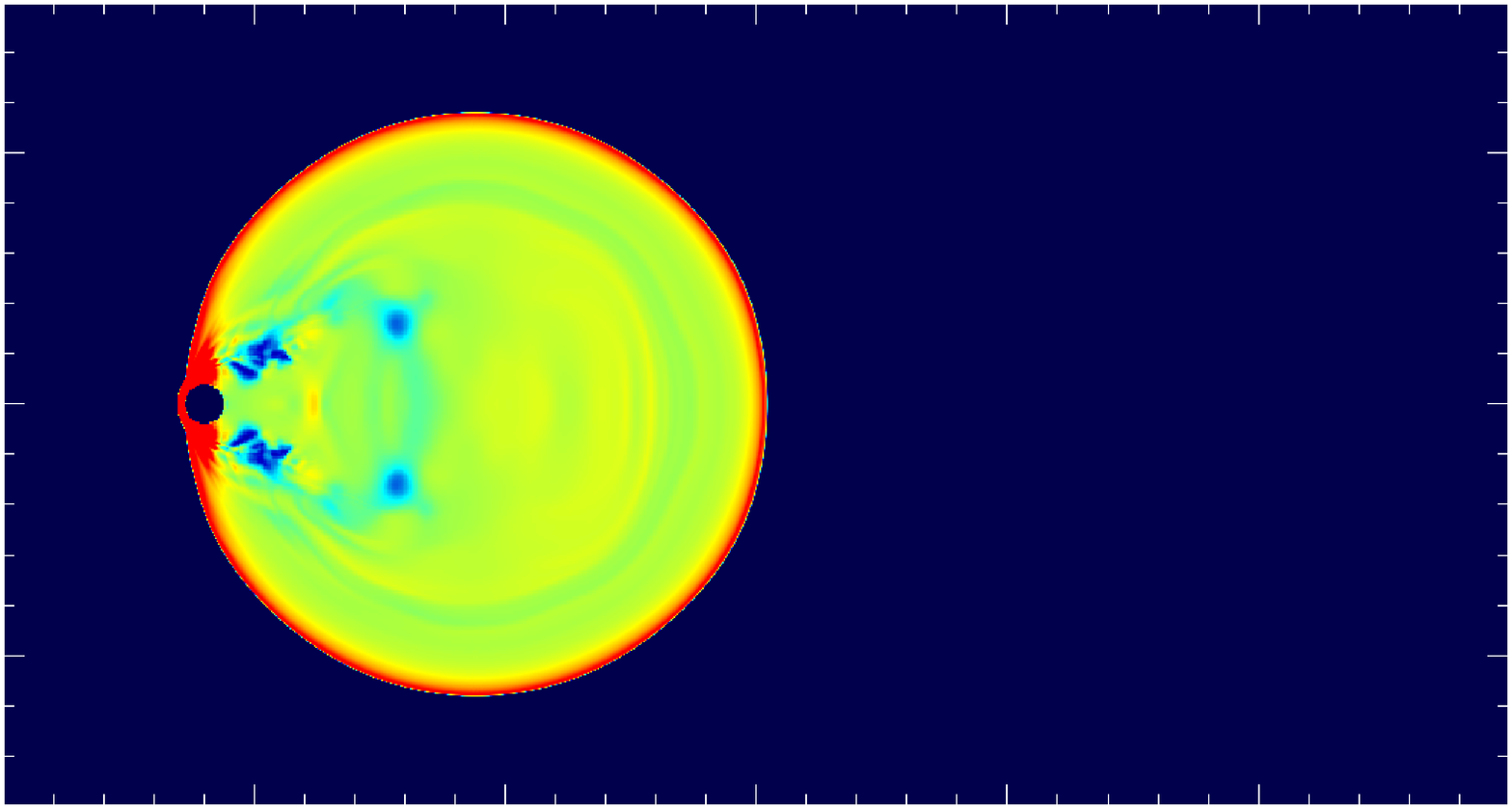}}
\subfloat[320,000 iterations.\label{fig:evo-e}]
{\includegraphics[angle=-90,width=1.8in]{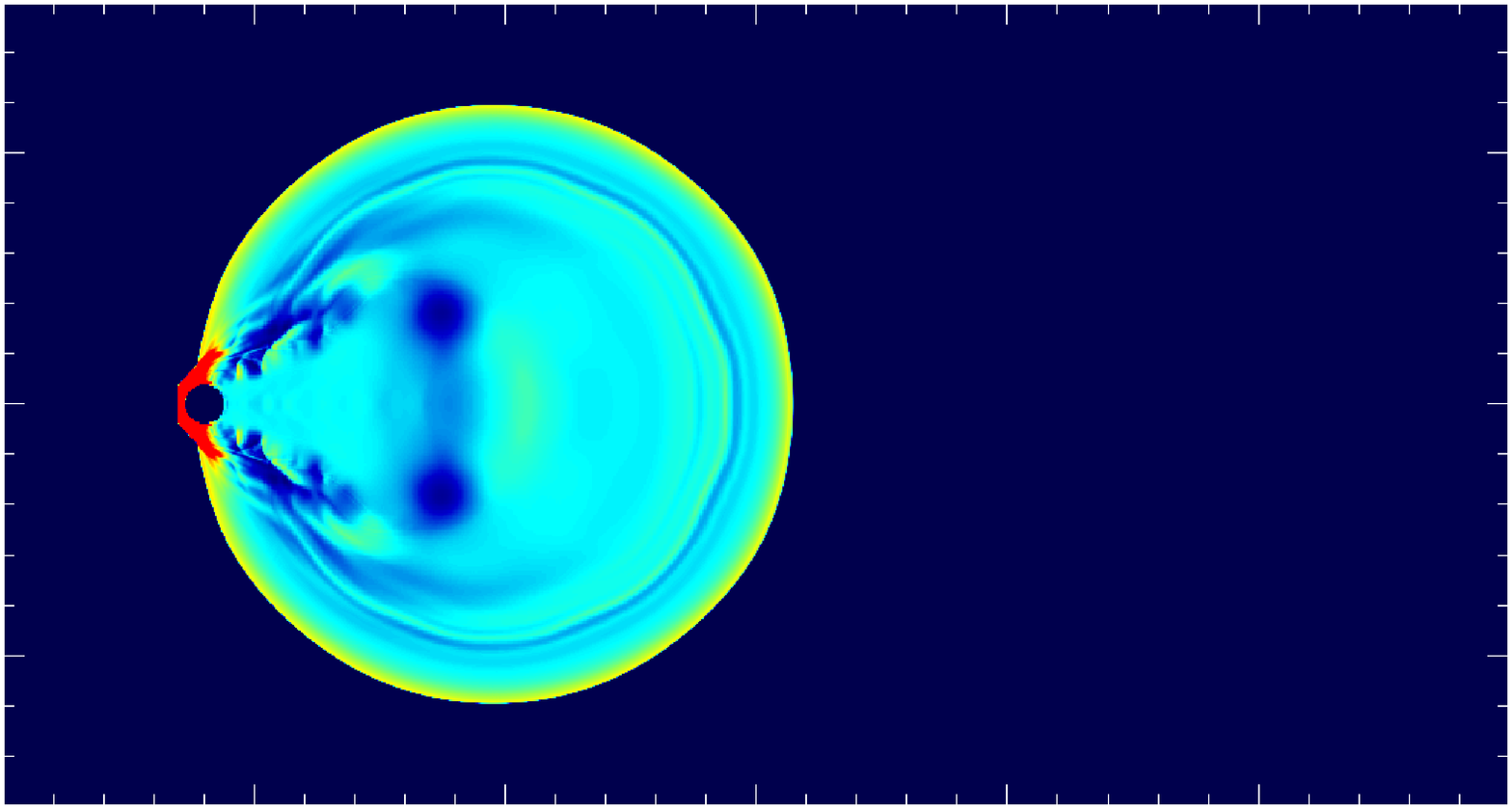}}
\subfloat[360,000 iterations.\label{fig:evo-f}]
{\includegraphics[angle=-90,width=1.8in]{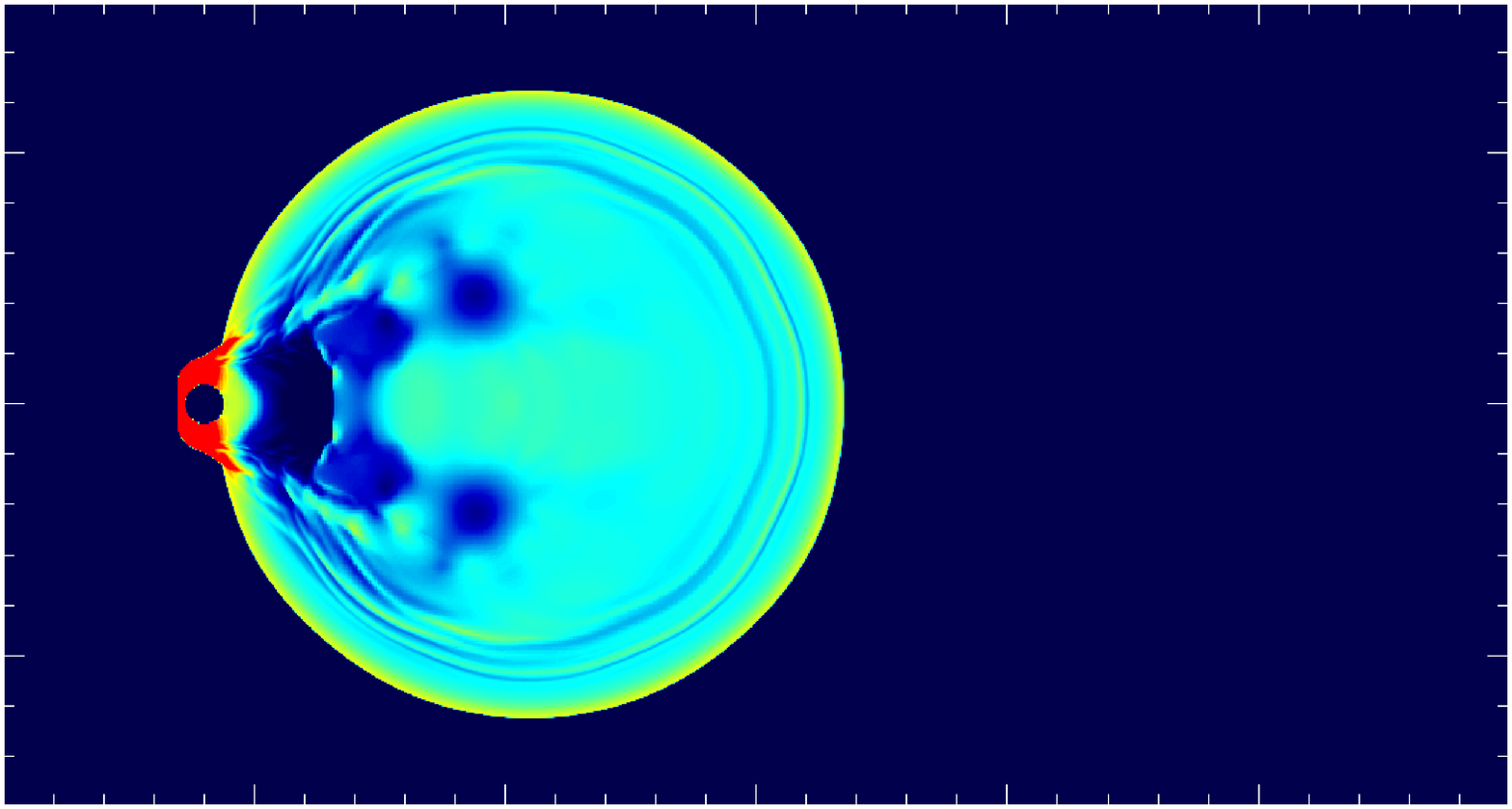}}\\
\subfloat[480,000 iterations.\label{fig:evo-g}]
{\includegraphics[angle=-90,width=1.8in]{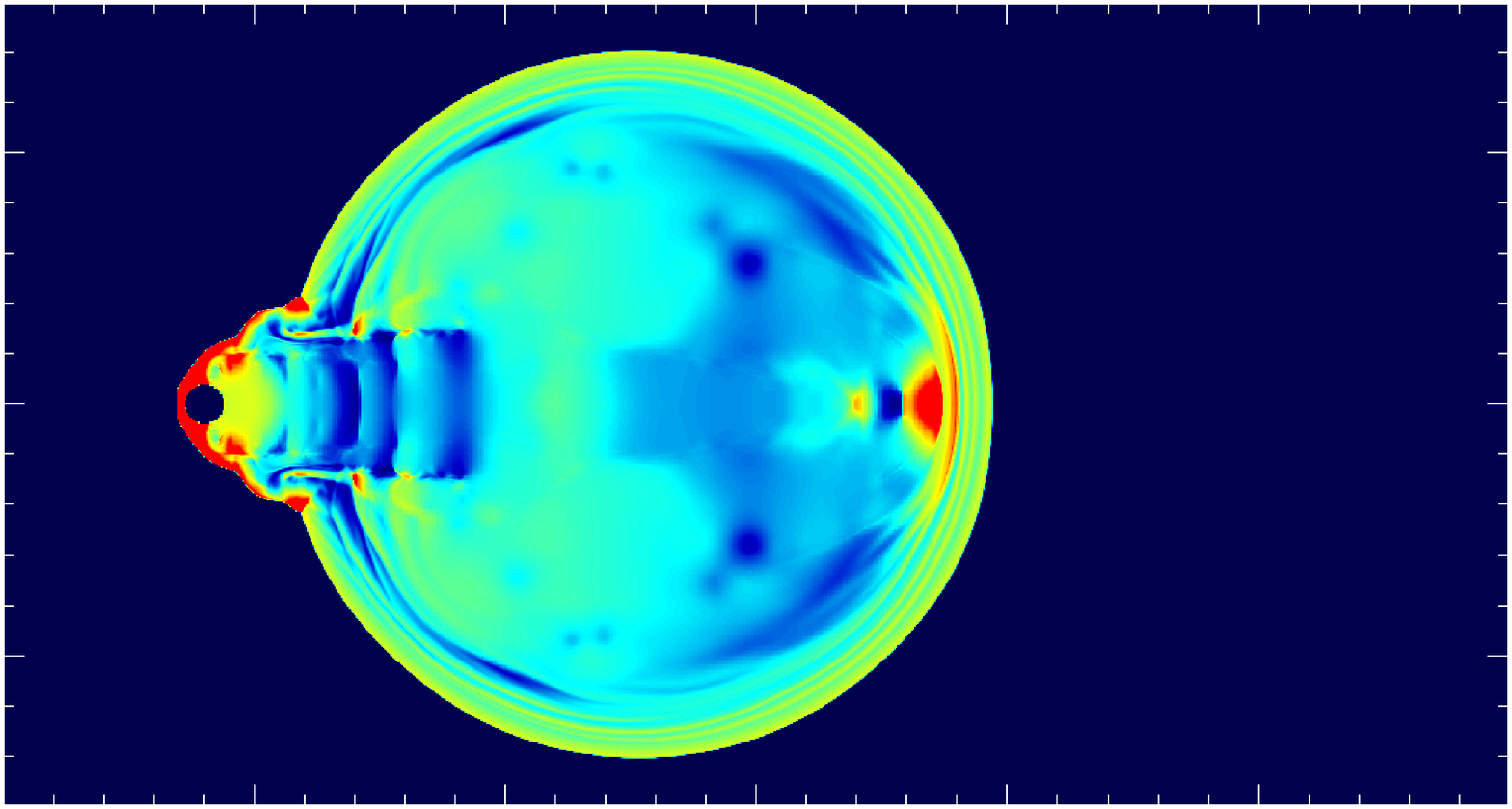}}
\subfloat[600,000 iterations.\label{fig:evo-h}]
{\includegraphics[angle=-90,width=1.8in]{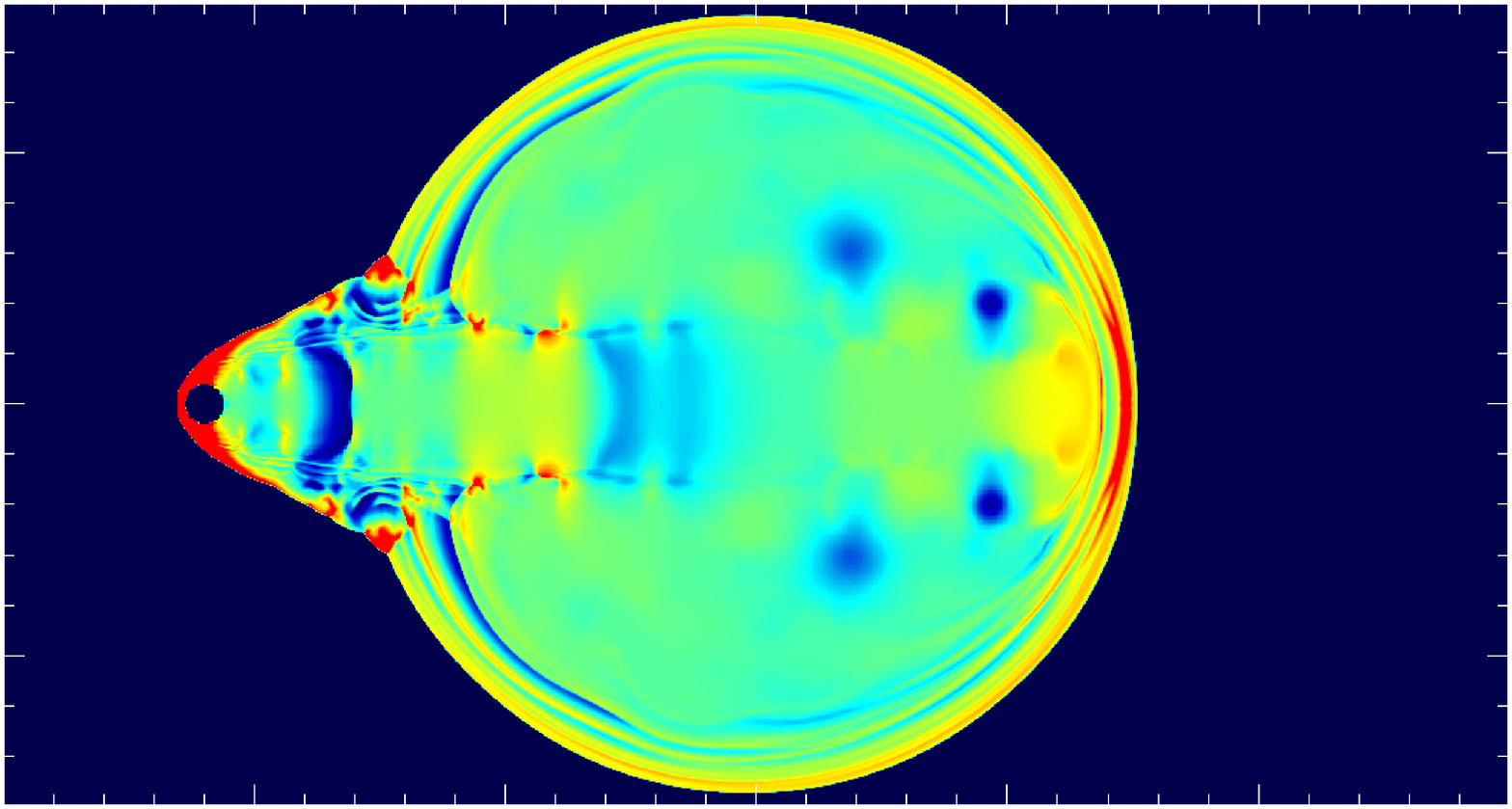}}
\subfloat[720,000 iterations.\label{fig:evo-i}]
{\includegraphics[angle=-90,width=1.8in]{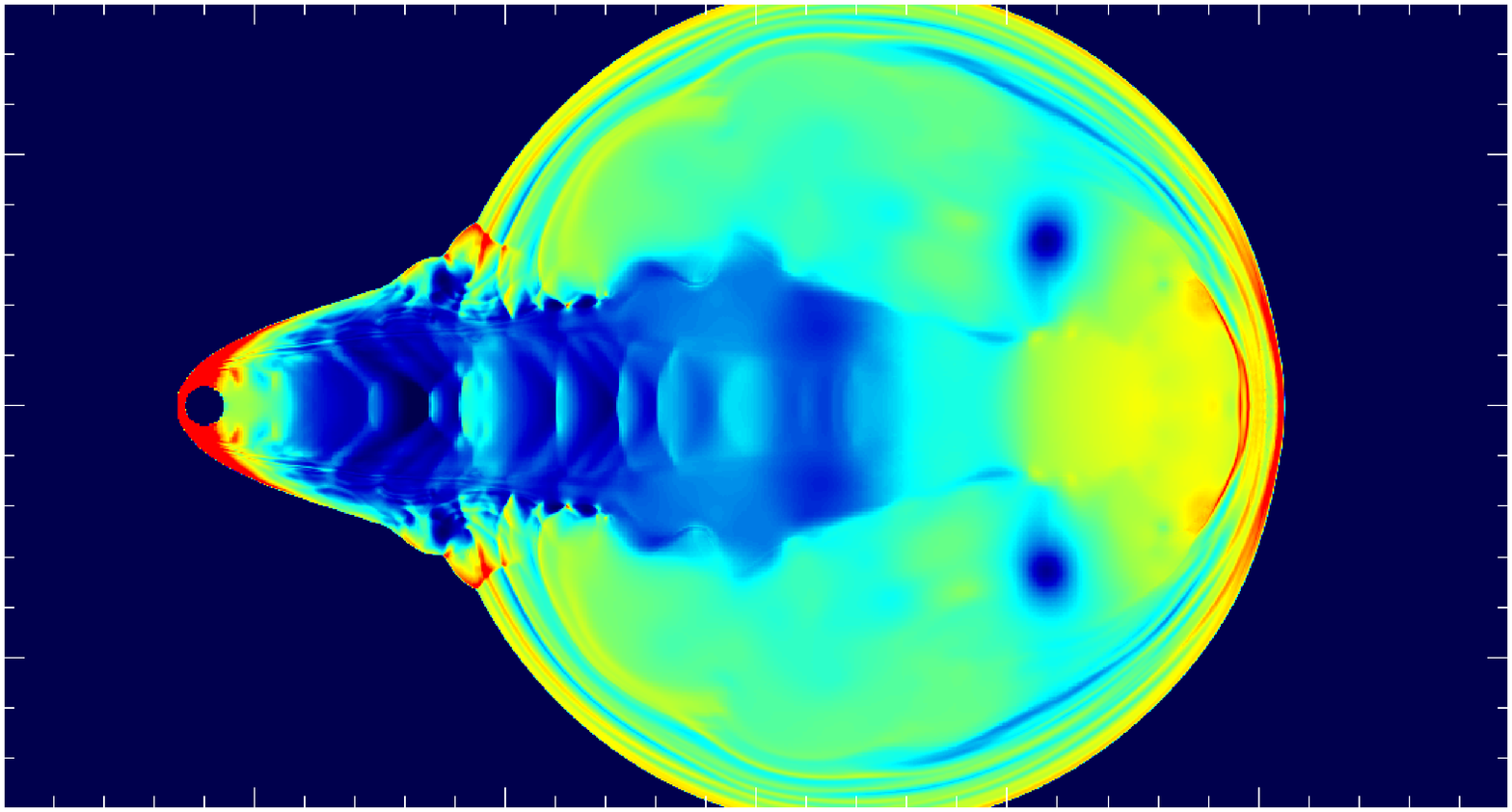}}\\[-10pt]
\label{fig:sim-evo}%
\end{figure}

\ber
\Delta v_{12} & = &\sqrt{\left(\frac{\cos\theta_1-\cos\theta_2}{1-\cos\theta_1\cos\theta_2}\right)^2+ \left(\frac{\sin\theta_1-\sin\theta_2}{1-\sin\theta_1\sin\theta_2}\right)^2}\nonumber\\
&\approx & 0.80\rm,\nonumber\\
\Delta v_{34} & = &\sqrt{\left(\frac{\cos\theta_3-\cos\theta_4}{1-\cos\theta_3\cos\theta_4}\right)^2+ \left(\frac{\sin\theta_3-\sin\theta_4}{1-\sin\theta_3\sin\theta_4}\right)^2}\nonumber\\
&\approx & 0.03\rm.\nonumber
\eer
This shows that the relative velocity between vertically adjacent on-axis cells just outside the hemisphere is supersonic relative to the pulsar outflow sound speed of 0.57 (for the parameters relevant to Fig.~\ref{sim}). Thus, the wind near the axis shocks immediately and is thermalized producing a post-termination shock flow. Given that at early times the wind shows no deviation from spherical symmetry (see Fig.~\ref{fig:sim-evo}), it is clear that this asymmetric numerical shocking of the wind is smeared out by the interaction with the ambient flow and does not impact the global evolution of the simulation.

Additional refinement levels, perhaps needed only at early simulation times, will mitigate the numerical shocking issue. However, since tests have shown such shocking is present with 3 levels, and the significant results discussed below were possible with 2 (i.e. L\tsb{max} = 1), we leave explorations of additional refinement levels to future studies. When these studies result in unshocked, ultra-relativistic wind flows entering the computational domain, and the resolution of the termination shock, we will perform new shock-tube tests. However, while the refined REST\_FRAME routine is essential for proper handling of the $\gamma >> 1$ outflow, in the simulations presented here there are no structures involving Lorentz factors higher than those previously explored by \cite{dun94} \& \cite{hug02} with the RHLLE solver for which shock-tube validation was performed. All structures referenced below originated in the computational domain where the the tried-and-true Newton-Raphson iterative solver was toggled into action (recall $\S$\ref{solver}). Therefore, we procede with firm confidence rooted in the previous shock-tube tests. 

\section{Discussion}\label{sec:dis}
The physics behind the formation of the structure observed in Fig.~\ref{sim} is as follows. The wind
streams outward and sweeps up ambient material which drives pressure
waves (weakly at first) into the shocked wind. As the nebula expands, the pressure inside decreases. Fig.~\ref{fig:sim-evo} shows a series of pressure maps comprising a time development sequence for the simulation depicted in Fig.~\ref{sim}. The sequence clearly indicates that once the pulsar crosses the boundary of the initially spherical nebula, an inflection point develops along the leading edge at approximately 45$^{\circ}$ from the axis as measured from W to N\footnote{This is sensible as it is the location where the wind velocity transitions from having its largest component at 180$^{\circ}$ to the inflow direction to having it at 90$^{\circ}$.}. 
The pressure waves are intensified and propagate to the axis and reflect, leading to the formation of a relativistic backflow harboring internal shockwaves reminiscent of shock diamonds. The fact that the backflow does not develop until after the inflection point supports this picture. The internal shockwaves, in turn, thermalize energy, allowing the flow to expand and inflate the
trailing spherical bubble. As the bubble inflates, it ``pinches'' the inflection point enhancing the cuspy shape, maintaining the pressure-wave influx that sets up the energy-thermalizing backflow responsible for inflating the bubble. Such a feedback cycle is relevent to the Guitar nebula even though the pulsar was not born at the center of the trailing bubble -- given its proper motion, the pulsar moves a distance corresponding to the entire nebula in less than 500 yr \citep{rom97}, a time orders of magnitude too short for the age of a pulsar powering a bow-shock nebula -- because it explains how the bubble persists. Such a scenario is analogous to the formation of structure in relativistic galactic jets, where the evolution is driven by Kelvin-Helmholtz modes long the contact surface that separates the shocked ambient medium from the shocked jet material \citep[e.g.][]{hug02}.

The evolution of the Guitar-like shape is rather sensitive to the choice of parameters. As Tab.~\ref{tab:it-inf} shows, the appearance of the inflection point marking the onset of the formation of the ``neck'' of the Guitar takes a significantly larger number of computational iterations as the ambient-flow velocity decreases. This is expected as the asymmetry of the nebula should evolve more slowly in this scenario: a decreased ambient-flow velocity is equivalent to a lower pulsar space motion and so it takes more time for the pulsar to reach the nebula boundary thereby delaying the formation of the inflection point. If the inflection point is not induced while the expanding nebula is small enough such that the ensuing neck is of significant scale, then no Guitar-like structure will be apparent. If a pulsar velocity of 1500 km~s$^{-1}$, 1250 km~s$^{-1}$, or 1000 km~s$^{-1}$ is required for observable Guitar-like morphology to arise, then the velocity distribution of \cite{arz02} implies that $<$5\%, 7-8\%, or $\sim$15\% of radio pulsars, respectively, have the possibility of developing such features depending on the nature of their ambient environment.

\linespread{1}
\begin{table}
\caption[The dependence of the Guitar-like inflection point on the number of simulation iterations.]{The dependence of the Guitar-like inflection point on the number of simulation iterations. As expected, the higher the ambient-flow velocity, the sooner the inflection point develops due to the increased rate at which ambient material is swept-up.}
\begin{center}
\begin{tabular}{ccc}
\hline\hline\relax\\[-1.7ex]
Wind Lorentz factor & ambient-flow velocity & Iterations until inflection\\
(unitless) & (km s$^{-1}$) & (10$^{4}$)\\\relax\\[-1.7ex]
\hline\relax\\[-1.7ex]
10$^5$ & &\\
       & 1750 & 30\\
       & 1500 & 39\\
       & 1250 & 54\\
       & 1000 & 81\\
       & 750  & unseen at 74\\
10$^4$ & & \\
       & 5500 & 4\\
       & 4250 & 6\\
       & 3000 & 12\\
       & 1750 & 30\\\relax\\[-1.7ex]
\hline
\label{tab:it-inf}
\end{tabular}
\end{center}
\end{table}

\section{Conclusion}\label{sec:conc}
We discussed the application of an existing special relativistic,
hydrodynamic (SRHD) primitive-variable recovery algorithm to
ultra-relativistic flows (Lorentz factor, $\gamma$, of
10$^2$--10$^6$) and the refinement necessary for the numerical
velocity root finder to work in this domain. 
We found that the velocity quartic, $Q(v)$,
exhibits dual roots in the physical velocity range that move
progressively closer together for larger $\gamma$ leading to a
divide by zero and the failure of the Newton-Raphson iteration
method employed by the existing primitives algorithm. Our solution
was to recast the quartic to be a function, $Q(\gamma)$, of $\gamma$.
We demonstrated that $Q(\gamma)$ exhibits only one physical root.
However, Newton-Raphson iteration also failed in this case at high
$\gamma$, due to the extreme slope of the quartic near the root,
necessitating the use there of an analytical numerical root finder.

Our timing analysis indicated that using $Q(\gamma)$ with the 8-byte
analytical root finder increased run time by only 24\% compared to using
$Q(v)$ with the 8-byte iterative root finder (based on 10 trial runs),
while using $Q(\gamma)$ with the 16-byte analytical root finder ballooned
run time by a factor of approximately 400. The iterative root finder is
accurate to order 10$^{-4}$ for a sizable region of parameter space
corresponding to Lorentz factors on the order of 10$^2$ and smaller.
Therefore, we implemented a computational switch that checks the values of $M/E$ and $R/E$ 
and calls the iterative or analytical root finder accordingly thereby creating a
hybrid primitives recovery algorithm called REST\_FRAME.

In addition, our exploration of parameter space suggests that the discriminant
of the cubic resolvent (as defined by Eqn.~\ref{disc} in $\S$\ref{solveqe}) will always be positive
for physical flows. Therefore, we did not include code for negative
discriminants in our routine. Formal proof remains elusive, however,
leaving potential for future work.

We have shown that REST\_FRAME is capable of calculating the
primitive variables from the conserved variables to an accuracy of
at least $O(10^{-4})$ for Lorentz factors up to 10$^6$ with
significantly better accuracy for Lorentz factors $\leq 10^5$, and
slightly worse (order 10$^{-3}$) for a small portion of the space
corresponding to the highest Lorentz factors. We traced the
degradation in accuracy for larger Lorentz factors to the effect of
subtractive cancellation. Past studies have shown that an accuracy
of order 10$^{-4}$ is capable of robustly capturing hydrodynamic
structures. We have applied the refined solver to an ultra-relativistic
problem and have shown that it is capable of reproducing observed
structures and is well-suited to our study of the
internal structure of diffuse pulsar wind nebulae.

Our main conclusions are as follows:
\begin{itemize}
\item Relativistic, hydrodynamic simulations have shown that the relatively slow, dense ISM flow resulting from the space motion of a pulsar can set up an interaction with the \textit{extremely} light, ultra-relativistic pulsar wind leading to an asymmetric nebula with a morphology reminiscent of the Guitar nebula. 
\item Simulations have validated the interpretation that a relativistic backflow behind PSR1929+10 is responsible for the X-ray morphology. Results further show that the backflow can harbor a series of internal shockwaves that inflates a nebular bubble, and that the bubble provides positive feedback to the backflow, explaining how the Guitar bubble persists.
\item The evolution of the bubble/backflow structure is sensitive to the choice of input parameters justifying a future series of simulation runs that will determine what pulsar velocities and wind/ISM density ratios are required for the bubble/backflow feedback loop to arise.
\end{itemize}
\section*{Acknowledgments}
This work was majority supported by NASA Graduate Student Researchers Program Grant $\#$NGT5-159. Steve Kuhlmann of Argonne National Laboratory provided access to additional computing resources beyond those used at the University of Michigan.


\begin{thebibliography}{48}
\expandafter\ifx\csname natexlab\endcsname\relax\def\natexlab#1{#1}\fi
\expandafter\ifx\csname url\endcsname\relax
  \def\url#1{\texttt{#1}}\fi
\expandafter\ifx\csname urlprefix\endcsname\relax\def\urlprefix{URL }\fi

\bibitem[{{Arzoumanian} et~al.(2002){Arzoumanian}, {Chernoff}, and
  {Cordes}}]{arz02}
{Arzoumanian}, Z., {Chernoff}, D.~F., {Cordes}, J.~M., Mar. 2002. {The Velocity
  Distribution of Isolated Radio Pulsars}. \apj 568, 289--301.

\bibitem[{{Bronshtein} and {Semendyayev}(1997)}]{bro97}
{Bronshtein}, I.~N., {Semendyayev}, K.~A., 1997. {Handbook of Mathematics, ed.
  K.~A. Hirsch}, 3rd Edition. Springer-Verlag Telos.

\bibitem[{{Bucciantini}(2008)}]{buc08}
{Bucciantini}, N., 2008. {Modeling Pulsar Wind Nebulae}. Advances in Space
  Research 41, 491--502.

\bibitem[{{Bucciantini} et~al.(2005){Bucciantini}, {Amato}, and {Del
  Zanna}}]{buc05}
{Bucciantini}, N., {Amato}, E., {Del Zanna}, L., Apr. 2005. {Relativistic MHD
  simulations of pulsar bow-shock nebulae}. \aap 434, 189--199.

\bibitem[{{Chatterjee} and {Cordes}(2002)}]{cha02}
{Chatterjee}, S., {Cordes}, J.~M., Aug. 2002. {Bow Shocks from Neutron Stars:
  Scaling Laws and Hubble Space Telescope Observations of the Guitar Nebula}.
  \apj 575, 407--418.

\bibitem[{{Cordes} and {Chernoff}(1998)}]{cor98}
{Cordes}, J.~M., {Chernoff}, D.~F., Sep. 1998. {Neutron Star Population
  Dynamics. II. Three-dimensional Space Velocities of Young Pulsars}. \apj 505,
  315--338.

\bibitem[{{Cordes} et~al.(1993){Cordes}, {Romani}, and {Lundgren}}]{cor93}
{Cordes}, J.~M., {Romani}, R.~W., {Lundgren}, S.~C., Mar. 1993. {The Guitar
  nebula - A bow shock from a slow-spin, high-velocity neutron star}. \nat 362,
  133--135.

\bibitem[{{Del Zanna} et~al.(2004){Del Zanna}, {Amato}, and
  {Bucciantini}}]{zan04}
{Del Zanna}, L., {Amato}, E., {Bucciantini}, N., Jul. 2004. {Axially symmetric
  relativistic MHD simulations of Pulsar Wind Nebulae in Supernova Remnants. On
  the origin of torus and jet-like features}. \aap 421, 1063--1073.

\bibitem[{{Del Zanna} and {Bucciantini}(2002)}]{zan02}
{Del Zanna}, L., {Bucciantini}, N., Aug. 2002. {An efficient shock-capturing
  central-type scheme for multidimensional relativistic flows. I.
  Hydrodynamics}. \aap 390, 1177--1186.

\bibitem[{{Del Zanna} et~al.(2003){Del Zanna}, {Bucciantini}, and
  {Londrillo}}]{zan03}
{Del Zanna}, L., {Bucciantini}, N., {Londrillo}, P., Mar. 2003. {An efficient
  shock-capturing central-type scheme for multidimensional relativistic flows.
  II. Magnetohydrodynamics}. \aap 400, 397--413.

\bibitem[{{Delettrez} et~al.(2005){Delettrez}, {Myatt}, {Radha}, {Stoeckl},
  {Skupsky}, and {Meyerhofer}}]{del05}
{Delettrez}, J.~A., {Myatt}, J., {Radha}, P.~B., {Stoeckl}, C., {Skupsky}, S.,
  {Meyerhofer}, D.~D., Dec. 2005. {Hydrodynamic simulations of integrated
  experiments planned for the OMEGA/OMEGA EP laser systems}. Plasma Physics and
  Controlled Fusion 47, B791--B798.

\bibitem[{{Duncan} and {Hughes}(1994)}]{dun94}
{Duncan}, G.~C., {Hughes}, P.~A., Dec. 1994. {Simulations of relativistic
  extragalactic jets}. \apjl 436, L119--L122.

\bibitem[{{Emmering} and {Chevalier}(1987)}]{emm87}
{Emmering}, R.~T., {Chevalier}, R.~A., Oct. 1987. {Shocked relativistic
  magnetohydrodynamic flows with application to pulsar winds}. \apj 321,
  334--348.

\bibitem[{{Fryxell} et~al.(2000){Fryxell}, {Olson}, {Ricker}, {Timmes},
  {Zingale}, {Lamb}, {MacNeice}, {Rosner}, {Truran}, and {Tufo}}]{fry00}
{Fryxell}, B., {Olson}, K., {Ricker}, P., {Timmes}, F.~X., {Zingale}, M.,
  {Lamb}, D.~Q., {MacNeice}, P., {Rosner}, R., {Truran}, J.~W., {Tufo}, H.,
  Nov. 2000. {FLASH: An Adaptive Mesh Hydrodynamics Code for Modeling
  Astrophysical Thermonuclear Flashes}. \apjs 131, 273--334.

\bibitem[{{Gaensler} and {Slane}(2006)}]{gae06}
{Gaensler}, B.~M., {Slane}, P.~O., Sep. 2006. {The Evolution and Structure of
  Pulsar Wind Nebulae}. \araa 44, 17--47.

\bibitem[{{Gammie} et~al.(2003){Gammie}, {McKinney}, and {T{\'o}th}}]{gam03}
{Gammie}, C.~F., {McKinney}, J.~C., {T{\'o}th}, G., May 2003. {HARM: A
  Numerical Scheme for General Relativistic Magnetohydrodynamics}. \apj 589,
  444--457.

\bibitem[{{Godunov}(1959)}]{god59}
{Godunov}, S.~K., 1959. {Difference Methods for the Numerical Calculations of
  Discontinuous Solutions of the Equations of Fluid Dynamics}. Mat. Sb. 47,
  271--306, in Russian, translation in: US Joint Publ. Res. Service, JPRS, 7226
  (1969).

\bibitem[{{Heger} et~al.(2003){Heger}, {Fryer}, {Woosley}, {Langer}, and
  {Hartmann}}]{hag03}
{Heger}, A., {Fryer}, C.~L., {Woosley}, S.~E., {Langer}, N., {Hartmann}, D.~H.,
  Jul. 2003. {How Massive Single Stars End Their Life}. \apj 591, 288--300.

\bibitem[{{Hirano}(2004)}]{hir04}
{Hirano}, T., Aug. 2004. {Hydrodynamic models}. Journal of Physics G Nuclear
  Physics 30, S845--S851.

\bibitem[{{Hughes}(2005)}]{hug05}
{Hughes}, P.~A., Mar. 2005. {The Origin of Complex Behavior of Linearly
  Polarized Components in Parsec-Scale Jets}. \apj 621, 635--642.

\bibitem[{{Hughes} et~al.(2002){Hughes}, {Miller}, and {Duncan}}]{hug02}
{Hughes}, P.~A., {Miller}, M.~A., {Duncan}, G.~C., Jun. 2002.
  {Three-dimensional Hydrodynamic Simulations of Relativistic Extragalactic
  Jets}. \apj 572, 713--728.

\bibitem[{{Ibanez}(2003)}]{iba03}
{Ibanez}, J.~M., 2003. {Numerical Relativistic Hydrodynamics}. In:
  {Fern{\'a}ndez-Jambrina}, L., {Gonz{\'a}lez-Romero}, L.~M. (Eds.), Current
  Trends in Relativistic Astrophysics. Vol. 617 of Lecture Notes in Physics,
  Berlin Springer Verlag. pp. 113--129.

\bibitem[{{Kargaltsev} and {Pavlov}(2008)}]{kar08}
{Kargaltsev}, O., {Pavlov}, G.~G., Jan. 2008. {Pulsar Wind Nebulae in the
  Chandra Era}. ArXiv e-prints 801. Available from: $<$http://adsabs.harvard.edu/
     abs/2008AIPC..983..171K$>$.


\bibitem[{{Kennel} and {Coroniti}(1984{\natexlab{a}})}]{kc84a}
{Kennel}, C.~F., {Coroniti}, F.~V., Aug. 1984{\natexlab{a}}. {Confinement of
  the Crab pulsar's wind by its supernova remnant}. \apj 283, 694--709.

\bibitem[{{Kennel} and {Coroniti}(1984{\natexlab{b}})}]{kc84b}
{Kennel}, C.~F., {Coroniti}, F.~V., Aug. 1984{\natexlab{b}}.
  {Magnetohydrodynamic model of Crab nebula radiation}. \apj 283, 710--730.

\bibitem[{{Komissarov}(1999)}]{kom99}
{Komissarov}, S.~S., Feb. 1999. {A Godunov-type scheme for relativistic
  magnetohydrodynamics}. \mnras 303, 343--366.

\bibitem[{{Mart{\'{\i}}} and {M{\"u}ller}(2003)}]{mar03}
{Mart{\'{\i}}}, J.~M., {M{\"u}ller}, E., Dec. 2003. {Numerical Hydrodynamics in
  Special Relativity}. Living Reviews in Relativity 6.

\bibitem[{{Michel}(1969)}]{mic69}
{Michel}, F.~C., Nov. 1969. {Relativistic Stellar-Wind Torques}. \apj 158,
  727--738.

\bibitem[{{Michel}(1973)}]{mic73}
{Michel}, F.~C., Mar. 1973. {Rotating Magnetospheres: an Exact 3-D Solution}.
  \apjl 180, L133--L137.

\bibitem[{{Mignone} and {McKinney}(2007)}]{mig07}
{Mignone}, A., {McKinney}, J.~C., Jul. 2007. {Equation of state in relativistic
  magnetohydrodynamics: variable versus constant adiabatic index}. \mnras 378,
  1118--1130.

\bibitem[{{Noble}(2003)}]{nob03}
{Noble}, S.~C., Oct. 2003. {A Numerical Study of Relativistic Fluid Collapse}.
  ArXiv General Relativity and Quantum Cosmology e-prints.

\bibitem[{{Noble} et~al.(2006){Noble}, {Gammie}, {McKinney}, and {Del
  Zanna}}]{nob06}
{Noble}, S.~C., {Gammie}, C.~F., {McKinney}, J.~C., {Del Zanna}, L., Apr. 2006.
  {Primitive Variable Solvers for Conservative General Relativistic
  Magnetohydrodynamics}. \apj 641, 626--637.

\bibitem[{{Pacini} and {Salvati}(1973)}]{pac73}
{Pacini}, F., {Salvati}, M., Nov. 1973. {On the Evolution of Supernova
  Remnants. Evolution of the Magnetic Field, Particles, Content, and
  Luminosity}. \apj 186, 249--266.

\bibitem[{{Perret-Gallix}(2006)}]{per06}
{Perret-Gallix}, D., 2006. {Concluding remarks: Emerging topics}. Proceedings
  of the X International Workshop on Advanced Computing and Analysis Techniques
  in Physics Research - ACAT 05.

\bibitem[{{Rees} and {Gunn}(1974)}]{rg74}
{Rees}, M.~J., {Gunn}, J.~E., Apr. 1974. {The origin of the magnetic field and
  relativistic particles in the Crab Nebula}. \mnras 167, 1--12.

\bibitem[{{Romani} et~al.(1997){Romani}, {Cordes}, and {Yadigaroglu}}]{rom97}
{Romani}, R.~W., {Cordes}, J.~M., {Yadigaroglu}, I.-A., Aug. 1997. {X-Ray
  Emission from the Guitar Nebula}. \apjl 484, L137--L140.

\bibitem[{{Ryu} et~al.(2006){Ryu}, {Chattopadhyay}, and {Choi}}]{ryu06}
{Ryu}, D., {Chattopadhyay}, I., {Choi}, E., Sep. 2006. {Equation of State in
  Numerical Relativistic Hydrodynamics}. \apjs 166, 410--420.

\bibitem[{{Schneider} et~al.(1993){Schneider}, {Katscher}, {Rischke},
  {Waldhauser}, {Maruhn}, and {Munz}}]{sch93}
{Schneider}, V., {Katscher}, U., {Rischke}, D.~H., {Waldhauser}, B., {Maruhn},
  J.~A., {Munz}, C.-D., Mar. 1993. {New Algorithms for Ultra-relativistic
  Numerical Hydrodynamics}. Journal of Computational Physics 105, 92--107.

\bibitem[{Shibata(2003)}]{shi03}
Shibata, M., Jan 2003. Axisymmetric general relativistic hydrodynamics:
  Long-term evolution of neutron stars and stellar collapse to neutron stars
  and black holes. Phys. Rev. D 67~(2), 024033.

\bibitem[{{Thompson}(1986)}]{tho86}
{Thompson}, K.~W., Oct. 1986. {The special relativistic shock tube}. Journal of
  Fluid Mechanics 171, 365--375.

\bibitem[{{van der Swaluw} et~al.(1998){van der Swaluw}, {Achterberg}, and
  {Gallant}}]{swa98}
{van der Swaluw}, E., {Achterberg}, A., {Gallant}, Y.~A., 1998. {Hydrodynamical
  simulations of pulsar wind nebulae in supernova remnants}. Memorie della
  Societa Astronomica Italiana 69, 1017--+.

\bibitem[{{van der Swaluw} et~al.(2003){van der Swaluw}, {Achterberg},
  {Gallant}, {Downes}, and {Keppens}}]{swa03}
{van der Swaluw}, E., {Achterberg}, A., {Gallant}, Y.~A., {Downes}, T.~P.,
  {Keppens}, R., Jan. 2003. {Interaction of high-velocity pulsars with
  supernova remnant shells}. \aap 397, 913--920.

\bibitem[{{van der Swaluw} et~al.(2001){van der Swaluw}, {Achterberg},
  {Gallant}, and {T{\'o}th}}]{swa01}
{van der Swaluw}, E., {Achterberg}, A., {Gallant}, Y.~A., {T{\'o}th}, G., Dec.
  2001. {Pulsar wind nebulae in supernova remnants. Spherically symmetric
  hydrodynamical simulations}. \aap 380, 309--317.

\bibitem[{{van der Swaluw} et~al.(2004){van der Swaluw}, {Downes}, and
  {Keegan}}]{swa04}
{van der Swaluw}, E., {Downes}, T.~P., {Keegan}, R., Jun. 2004. {An
  evolutionary model for pulsar-driven supernova remnants. A hydrodynamical
  model}. \aap 420, 937--944.

\bibitem[{{Vigelius} et~al.(2007){Vigelius}, {Melatos}, {Chatterjee},
  {Gaensler}, and {Ghavamian}}]{vig07}
{Vigelius}, M., {Melatos}, A., {Chatterjee}, S., {Gaensler}, B.~M.,
  {Ghavamian}, P., Jan. 2007. {Three-dimensional hydrodynamic simulations of
  asymmetric pulsar wind bow shocks}. \mnras 374, 793--808.

\bibitem[{{Wang} et~al.(1993){Wang}, {Li}, and {Begelman}}]{wan93}
{Wang}, Q.~D., {Li}, Z.-Y., {Begelman}, M.~C., Jul. 1993. {The X-ray-emitting
  trail of the nearby pulsar PSR1929 + 10}. \nat 364, 127--129.

\bibitem[{{Weaver} et~al.(1977){Weaver}, {McCray}, {Castor}, {Shapiro}, and
  {Moore}}]{wea77}
{Weaver}, R., {McCray}, R., {Castor}, J., {Shapiro}, P., {Moore}, R., Dec.
  1977. {Interstellar bubbles. II - Structure and evolution}. \apj 218,
  377--395.

\bibitem[{{Zhang} et~al.(2003){Zhang}, {Woosley}, and {MacFadyen}}]{zha03}
{Zhang}, W., {Woosley}, S.~E., {MacFadyen}, A.~I., Mar. 2003. {Relativistic
  Jets in Collapsars}. \apj 586, 356--371.

\end{thebibliography}
\end{document}